\pdfoutput=1
\documentclass[manuscript]{acmart}
\usepackage{xspace}
\usepackage{lscape} 
\usepackage{geometry}
\usepackage{pbox}
\usepackage{graphicx}
\usepackage{multirow}
\usepackage{amsthm}
\usepackage{amsmath}
\usepackage{color}
\usepackage{enumitem}
\usepackage{subcaption}
\usepackage{caption}
\usepackage{array}
\usepackage{booktabs}
\usepackage{makecell}
\usepackage{footnote}  
\usepackage{graphicx}
\usepackage{rotating}
\usepackage{comment}
\usepackage{todonotes}
\usepackage{tikz}
\usepackage{pgfplots}
\pgfplotsset{every tick label/.append style={font=\small}}
\usepackage{algorithmicx}
\usepackage{algorithm}
\usepackage[noend]{algpseudocode}
\usepackage{arydshln}

\algrenewcommand\algorithmicrequire{\textbf{Input:}}
\algrenewcommand\algorithmicensure{\textbf{Output:}}

\newcommand{\eps}{\varepsilon}
\newcolumntype{C}[1]{>{\centering\let\newline\\\arraybackslash\hspace{0pt}}m{#1}}

\newcommand{\enet}{{$\varepsilon$-net}\xspace}
\newcommand{\indep}{\perp \!\!\! \perp}

\AtBeginDocument{
  \providecommand\BibTeX{{
    \normalfont B\kern-0.5em{\scshape i\kern-0.25em b}\kern-0.8em\TeX}}}

\setcopyright{acmcopyright}
\copyrightyear{2021}
\acmYear{2021}
\acmDOI{10.1145/1122445.1122456}
\acmConference[Woodstock '18]{Woodstock '18: ACM Symposium on Neural
  Gaze Detection}{June 03--05, 2018}{Woodstock, NY}
\acmBooktitle{Woodstock '18: ACM Symposium on Neural Gaze Detection,
  June 03--05, 2018, Woodstock, NY}
\acmPrice{15.00}
\acmISBN{978-1-4503-XXXX-X/18/06}

\newtheorem{example}{Example}

\begin{document}

\title{Representation Bias in Data: A Survey on Identification and Resolution Techniques}

\author{Nima Shahbazi}
\email{nshahb3@uic.edu}
\affiliation{
\institution{University of Illinois Chicago}
\country{USA}
}
\author{Yin Lin}
\email{irenelin@umich.edu }
\affiliation{
\institution{University of Michigan}
\country{USA}
}
\author{Abolfazl Asudeh}
\email{asudeh@uic.edu}
\affiliation{
\institution{University of Illinois Chicago}
\country{USA}
}
\author{H. V. Jagadish}
\email{jag@umich.edu }
\affiliation{
\institution{University of Michigan}
\country{USA}
}

\newcommand\nima[1]{\textcolor{purple}{(Nima) #1} }
\newcommand\yin[1]{\textcolor{orange}{(Yin) #1} }
\newcommand\abol[1]{\textcolor{red}{(Abol) #1} }
\newcommand\jag[1]{\textcolor{red}{(Jag) #1} }
\newcommand{\gee}{\mathcal{G}}
\newcommand{\dee}{\mathcal{D}}

\newcommand\new[1]{\textcolor{blue}{#1}\xspace}
\begin{abstract}
Data-driven algorithms are only as good as the data they work with, while data sets, especially social data, often fail to represent minorities adequately. Representation Bias in data can happen due to various reasons ranging from historical discrimination to selection and sampling biases in the data acquisition and preparation methods.
Given that ``bias in, bias out'', one cannot expect AI-based solutions to have equitable outcomes for societal applications, without addressing issues such as representation bias.
While there has been extensive study of fairness in machine learning models, including several review papers, bias in the data has been less studied.
This paper reviews the literature on identifying and resolving representation bias as a feature of a data set, independent of how consumed later.
The scope of this survey is bounded to structured (tabular) and unstructured (e.g., image, text, graph) data.
It presents taxonomies to categorize the studied techniques based on multiple design dimensions and provides a side-by-side comparison of their properties.

There is still a long way to fully address representation bias issues in data.
The authors hope that this survey motivates researchers to approach these challenges in the future by observing existing work within their respective domains.
\end{abstract}



\maketitle

\section{Introduction}
Data-driven decision-making shapes every corner of human life, from autonomous vehicles to healthcare and even predictive policing and criminal sentencing. A critical question, particularly in applications impacting human beings, is how trustworthy the decision made by the system is.
It is easy to see that the accuracy of a data-driven decision depends, first and foremost, on the data used to make it. After all, the system learns the phenomena that data represent. As a first step, we may desire that the data should represent the underlying data distribution from which the production data will be drawn. But that is not enough since it only tells us about the overall model performance. 
Although a system may generally perform well in terms of accuracy, it could fail for less populated regions in the data with insufficient representation. These regions may matter because they frequently represent some minority (sub)population in society. They could also represent cases that may not happen very often but have a relevant impact on the correctness of a critical decision.
In short, if data is not representative of a given population, the outcome of the decision system for that subpopulation may not be trustworthy.

{\em Representation Bias} happens when the training data under-represents (and subsequently fails to generalize well) some parts of the target population \cite{suresh2021framework}. 
Data representation bias can originate from how (and from where) the data was originally collected or be caused by the biases introduced after collection, either historically, cognitively, or statistically.
Representation bias can happen due to selection bias, i.e. when the sampling method only reaches a portion of the population or the population of interest has changed or is distinct from the population used during model training. For example, a survey to measure the illegal drug use of teenagers could be biased if it only includes high school students and ignores home-schooled students or dropouts. Another potential reason is the skewness of the underlying distribution. Suppose the target population for a particular medical data set is adults aged 18-60. There are minority groups within this population: for example, pregnant people may make up only 5\% of the target population. Even with perfect sampling and an identical population, the model is prone to be less robust for the group of pregnant people because it has fewer data points to learn from \cite{suresh2021framework}.
Furthermore, even if we carefully arrange for uniform sampling by age, we may find that sampling is non-uniform for pregnant people. For example, there may be proportionately fewer pregnant people over 40. If some group is a minority in the underlying distribution, then even random sampling will not help the under-representation issue for this group.

Representation bias is almost always guaranteed without a systematic approach to data collection. 
For example, in a survey data collection, 
a crucial step is to identify all the sub-populations in the underlying distribution based on the desired demographic information and ensure that the survey reaches all of them while enough samples are collected from each. However, the problem is that data scientists usually do not have any control over the data collection process, resulting in the utilization of ``found data'' in most data-driven decision-making systems. Therefore, with no guarantee on the aforementioned steps in the data collection process, the found data is most likely a biased sample. 

Representation bias in data is not a new problem and has been a known issue in data mining, database management, and statistics communities.
There is a rich line of work on the problem of discovering interesting patterns, regularities, or finding empty space in the data that is a parallel and relatively similar problem to identifying representation bias in data sets~\cite{10.5555/1622270.1622290, DBLP:conf/pricai/LiuWMQ98,DBLP:journals/tcs/EdmondsGLM03,DBLP:journals/corr/LemleyJA17}.
However, with the emergence of responsible data science and trustworthy AI, this problem has been addressed with greater vigor and from a brand new perspective in recent years. This survey discusses techniques for identifying and resolving representation bias in data sets, introducing taxonomies to classify these techniques based on multiple dimensions. Note that while the literature on algorithmic fairness is primarily concerned with promoting fairness in machine learning (ML) {\em models}, bias is sought to be addressed in the {\em data sets}, regardless of how the data is ultimately consumed.

We start the paper by presenting a big-picture overview of the {\em fairness literature} in Section~\ref{sec:fairness}. This will help us specify the scope of this survey w.r.t. fairness approaches and existing surveys.
Next, in Section~\ref{sec:overview}, we zoom in on the notion of {\em representation bias}, explaining the reasons that give rise to it, and presenting techniques for measuring representation bias.
In Section~\ref{sec:structured}, we propose a taxonomy to categorize different approaches to identify and resolve representation bias in {\em structured data} based on factors such as objectives and capabilities. 
Following our taxonomy's guidelines, we investigate each work's details, explain its novelty, and discuss its pros and cons. In Section~\ref{sec:unstructured}, we review the techniques for identifying and resolving representation bias in {\em unstructured data} such as images, text, speech, and graphs.
Finally, in Section~\ref{sec:conclusion}, we present an overview of the reviewed works and conclude the survey by discussing aspects that have been less noticed in the existing lines of work and propose some possible directions for the researchers to investigate.

\section{An Overview of Fairness Literature} \label{sec:fairness}
As AI replaces human beings in various critical fields, the topic of fairness among the affected population becomes more crucial. In recent years, the general topic of fairness has drawn sizable attention from different communities, specifically in the ML field. Many surveys \cite{pessach2022review,mehrabi2021survey,stoyanovich2020responsible,balayn2021managing, catania2022fairness} and tutorials have been published on the related topics and even the conference ACM FAccT \footnote{\url{https://facctconference.org/}} has been dedicated to this topic. 
Before focusing on representation bias in a data set, it is beneficial to review the big picture of fairness literature, including the definitions and techniques to achieve fairness. 
Given this context, we will specify the scope of this survey. 

\subsection{Definitions of Fairness}
There is no clear agreement on the definitions of fairness since it all depends on the task we target to solve and the numerous kinds of bias that can exist in data. However, at a high level, fairness definitions can be viewed from three perspectives ~\cite{fairmlbook}: \textit{individual fairness}, \textit{group fairness}, \textit{subgroup fairness}. 

\paragraph{\textbf{Individual Fairness}} Individual fairness is the most granular notion of fairness, requiring similar outcomes for similar individuals ~\cite{Dwork2012FairnessTA}.

\paragraph{\textbf{Group Fairness}}
Group fairness is the most popular category of fairness definitions for learning models. The term ``group'' refers to the classification of individuals within a population into a particular social category that has been historically subject to discriminatory treatment \cite{fairmlbook}. Examples of such social categories a.k.a. {\bf sensitive attributes} include {\em race, gender, sexual orientation, age, religion, disability}, etc.
A model satisfies some group fairness definition if it has equal or similar performance on different groups w.r.t. the associated fairness measures.
Most of ML group fairness metrics could be classified into the following categories ~\cite{fairmlbook,asudeh2020fairly}: {\em independence}, {\em separation}, {\em sufficiency}, {\em causation}. 

\textit{Independence} only relies on the model's predicted outcome, and a model satisfies independence if its outcome is independent of the sensitive attributes.
Let $h(x)$ and $\gee$ represent the model outcome and the demographic groups, respectively.
Under Independence measures\footnote{$\indep$ is the mathematical independence operation between two random variables.},

\vspace{-7mm}
\begin{align}\label{eq:def:independence}
h(x)\indep \gee
\end{align}
Measures such as {\em Statistical Parity}~\cite{Dwork2012FairnessTA} fall under this category. These measures indicate that different demographic groups have (almost) equal probabilities to generate positive (favorable) prediction: $\forall g_i, g_j \in \gee, Pr(h(x)=1\vert g_i) \simeq	 Pr(h(x)=1\vert g_j)$. 
{\em Conditional statistical parity}~\cite{CorbettDavies2017AlgorithmicDM} extends the definition of independence by considering a set of legitimate attributes $L$ that could affect the outcome: $\forall g_i, g_j \in \gee, Pr(h(x)=1\vert x_l = l, g_i) \simeq	 Pr(h(x)=1\vert x_l = l, g_j)$.
For example, suppose the demographic groups are {\tt male} and {\tt female}, and the legitimate factor is marital status. Therefore, the probability of {\tt married male} and {\tt married female} getting a positive prediction result should be equivalent.

\textit{Separation} is satisfied when the outcome of the model is independent of the sensitive attribute(s) conditioned on the ground-truth label $y$. That is,

\vspace{-9mm}
\begin{align}\label{eq:def:separation}
\big(h(x)\indep \gee\big)~\big|~y
\end{align}
Two well-known measures in this category are {\em Equalized Odds} and {\em Equal Opportunity}~\cite{hardt2016equality}.
Equalized odds is considered in contexts that correctly predicting positive outcomes and minimizing costly false positives are both of high importance: $\forall g_i\in \gee, Pr(h(x)=1 \vert g_i, y=1) \simeq Pr(h(x)=1\vert  y=1)$ and $Pr(h(x)=1 \vert g_i, y=0) \simeq Pr(h(x)=1 \vert  y=0)$.
Equal opportunity is a reasonable measure when predicting the positive outcome correctly is crucial and false positives are not costly: $\forall g_i\in \gee,\, Pr(h(x) = 1\vert g_i, y=1) \simeq Pr(h(x) = 1 \vert y=1)$.

\textit{Sufficiency}, on the other hand, is satisfied if, under the same model outcomes, sensitive attribute(s) and the true outcome are independent. That is,

\vspace{-9mm}
\begin{align}\label{eq:def:sufficiency}
\big(h(x)\indep y\big)~\big|~\gee
\end{align}
Sufficiency can be measured with \textit{Predictive Parity}~\cite{Chouldechova2017FairPW}. 
Positive predictive parity guarantees an equal chance of success, given the positive prediction for all subgroups:
$\forall g_i\in \gee, Pr(y=1 \vert h(x)=1 , g_i) \simeq Pr(y=1 \vert h(x)=1 )$.
Similarly, negative predictive parity ensures an equal chance of success given the negative prediction for all subgroups.

\textit{Causation}, aka counterfactual fairness \cite{kusner2017counterfactual,salimi2019interventional} focuses on the causal relationship between attributes, for instance when an attribute $A$ affects attribute $B$, which in turn affects attribute $C$. 
The counterfactual definition of fairness follows the intuition that a decision is fair for an individual if, in a counterfactual world, the decision would not change had the individual belonged to a different demographic group.

Please note that this is not an exhaustive list of group fairness definitions, and we only introduced the ones more commonly known and practiced. For a more exhaustive list and extensive discussion, please see \cite{fairmlbook,Verma2018FairnessDE}.

\paragraph{\textbf{Subgroup Fairness}}
Falling in between individual and group fairness, subgroup fairness~\cite{kearns2019empirical} (also known as intersectional fairness) metrics measure fairness (according to the above definitions) when groups are defined over the intersection of values of multiple sensitive attributes (e.g. {\tt white~male}, {\tt white~female}, {\tt black~male}, and {\tt black~female}).

Having discussed the existence of unfairness in ML models with the assistance of fairness definitions, next, we introduce strategies to promote fairness.

\subsection{Interventions to Achieve Fairness}
\begin{figure}
    \centering
    \includegraphics[width=\textwidth]{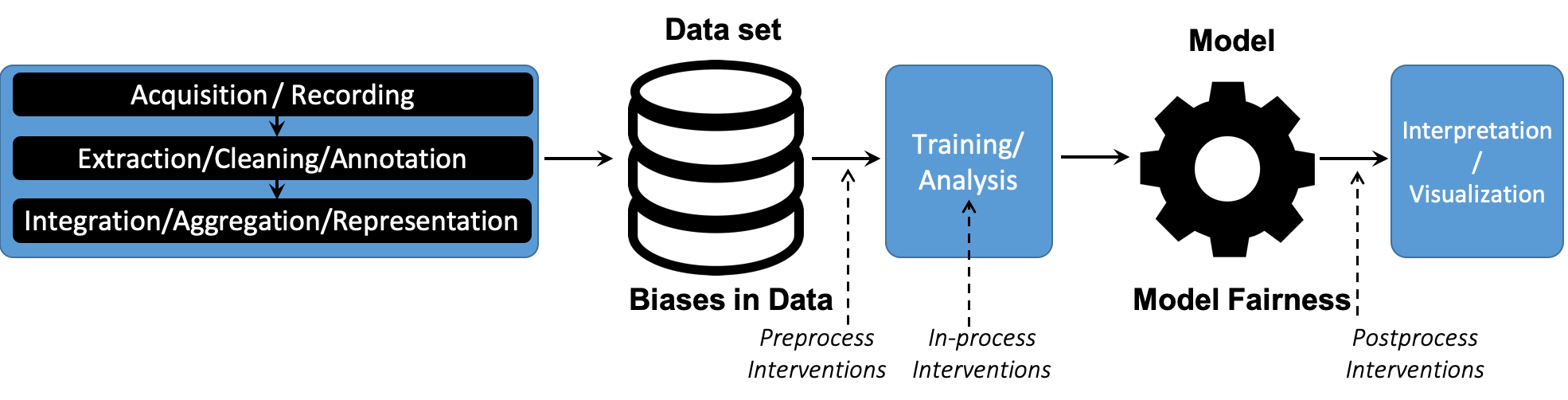}
    \caption{Illustration of bias and fairness in data analytics pipeline (the pipeline is adapted from \cite{jagadish2014big})}
    \label{fig:data-pipeline}
\end{figure}

Fairness can be considered by ML models~\cite{dAlessandro2019ADS,caton2020fairness} at different stages of the data analysis pipeline, shown in Figure~\ref{fig:data-pipeline}.
As highlighted in the figure, the intervention strategies to achieve model fairness fall under three categories: 
\textit{Pre-process}, \textit{In-process}, and \textit{Post-process} interventions. 

\paragraph{\textbf{Pre-process interventions}}
The main idea of this category of techniques is to modify the data before feeding it into the ML algorithms.
The common pre-process interventions include: \textit{data massaging}, \textit{reweighting}, \textit{sampling}, \textit{modifying feature representations}, \textit{adversarial learning}, and \textit{causal methods}. 

\textit{Data massaging}, first proposed by Kamiran et al. ~\cite{Kamiran2009ClassifyingWD}, aims to select the best candidates in the training data for relabeling by ranking the candidates according to their probability of belonging to the opposite class using a Naive Bayesian classifier. 

\textit{Data reweighting}~\cite{Calders2009BuildingCW} carefully assigns the tuples in the training set with different weights such that the new distribution is discrimination free with respect to the sensitive attributes. 

\textit{Sampling} methods \cite{Kamiran2011DataPT} can be used to under or over-sample the training data set for the ML algorithms that cannot directly work with weight. 
Given a sensitive attribute and considering the attribute value and label selection, there are four groups: two need over-sampling, and the other two need under-sampling. The employed sampling techniques include \textit{uniform sampling}, which applies uniform probability to increase or decrease the size of the groups, and \textit{preferential sampling}, where borderline objects get higher priority to be duplicated or ignored. 

\textit{Modifying feature representations} includes learning an intermediate representation that maintains all the essential information while removing any sign of the sensitive attribute. In \cite{Dwork2012FairnessTA}, to handle individual fairness, Dwork et al. propose to find a mapping from individuals to an intermediate representation that minimizes the loss subject while satisfying the Lipschitz condition to guarantee that similar individuals are treated similarly. Further, to produce out-of-sample representations to handle unseen examples, Zemel et al. \cite{Zemel2013LearningFR} develop a learning approach to achieve both group and individual fairness for ML models. The primary purpose is to learn a set of intermediate representations that satisfy two goals: first, they encode the data as well as possible, and second, they should be blind to whether the individual is from the protected group. The authors design a learning objective considering statistical parity, prediction accuracy, and data loss. iFair \cite{Lahoti2019iFairLI} is a learning-to-rank algorithm that introduces a method for probabilistically mapping user records into a low-rank representation to achieve individual fairness. Compared to the other learning representation algorithms, it is agnostic to downstream machine learning algorithms and can handle a broader range of applications. In \cite{Lahoti2019OperationalizingIF}, Lahoti et al. propose a method to model information of equally deserving individuals as a fairness graph. Based on the fairness graph, the proposed method learns a Fair Representation (PFR) to capture both data-driven similarities between individuals and pairwise side-information. Compared to the previous works for resolving individual fairness, PFR avoids the most challenging part of eliciting a quantitative measure of similarity from human experts. Optimized pre-processing \cite{Calmon2017OptimizedDP} formulates an optimization problem to probabilistically transform the data to trade off discrimination control, data utility, and individual fairness.

\textit{Adversarial learning} is another approach to increase the amount of data for the sensitive groups to achieve group fairness \cite{adel2019one,Xu2018FairGANFG}. There are also generative models to enhance the training data set of fair classification. FairGAN \cite{Xu2018FairGANFG} uses a generative adversarial network (GAN) to generate synthetic data to enhance group fairness when the original data is limited. It considers data utility, data fairness, classification utility, and classification fairness as important requirements for the generated data.

\textit{Causal methods} uncover the causal relationships in the data and focus on the dependencies between sensitive attributes and attributes acting as a proxy \cite{salimi2019interventional,galhotra2017fairness,chrisrussell,glymour2019measuring}. In this regard, training data repairing strategies have been suggested, such as \cite{salimi2019interventional} by Salimi et al., to minimally modify the databases by \textit{remove, insert, update} operations based on the notion of conditional independence between outcome and the sensitive attributes.

\paragraph{\textbf{In-process interventions}}
In-processing methods mainly reinforce fairness by inducing constraints or adding regularization terms to the objective function of the learning algorithm \cite{Kamishima2011FairnessawareLT, Zafar2017FairnessCM,agarwal2018reductions,goh2016satisfying,bechavod2017penalizing,woodworth2017learning,calders2010three}. 
The enforced constraints ensure that the algorithm treats different subpopulations equally w.r.t. the specified fairness measures.
Other in-processing approaches include adversarial learning \cite{wadsworth2018achieving,zhang2018mitigating,beutel2017data,beutel2019putting,edwards2015censoring, celis2019improved,xu2019achieving}, re-weighing \cite{krasanakis2018adaptive,jiang2020identifying,zhang2021omnifair}, and bandits \cite{joseph2016fairness, joseph2018meritocratic,liu2017calibrated,ensign2018decision,gillen2018online} approaches. Adversarial approaches use fairness measures to provide feedback to the model by penalizing it if the sensitive attributes are predictable from any of the remaining attributes. This is usually achieved by subjecting the model to many constraints and formulating the problem as a multiple-constraint optimization problem.
In-process re-weighting approaches usually begin by learning an unweighted classifier on the data and then using the learned weights of the samples to retrain the classifier. Bandit-based approaches usually cannot define what it means to be fair, but they may be able to recognize it when it is observed. This is usually achieved through the notions of individual fairness.

\paragraph{\textbf{Post-process interventions}}
Post-processing methods manipulate the results of a classifier to promote fairness among different groups. Hardt et al. \cite{Hardt2016EqualityOO} propose a post-processing technique to guarantee equalized odds by formulating it as an optimization problem that finds the probabilities that can be used to change the output labels to remove discrimination from protected groups. Calibrated equalized odds by Pleiss et al. \cite{Pleiss2017OnFA} explores the relationship between calibration and error rates. They provide an algorithm that aims to effectively find the unique feasible solution to satisfy both by determining probabilities used to flip the output labels. Reject option classification by Kamiran et al. \cite{Kamiran2012DecisionTF} invokes a reject option and labels instances in deprived and favored groups by the posterior probability to reduce discrimination.

\subsection{Scope of the Survey}
Having discussed the intervention approaches to achieve model fairness, let us look at the pipeline of data analytics in Figure~\ref{fig:data-pipeline} again.
The model is often considered ``the product'' of the pipeline.  Indeed, works on fairness focus on the model.  However, the {\em data set} is also a product, of possible interest in its own right, in addition to its influence on model fairness.
Given a data set to train a model, fairness intervention techniques aim to build a model based on some fairness criteria.
On the other hand, this survey focuses on the other product of the pipeline, i.e., {data sets}, studying {\em bias as a feature of a data set}, independent of how it is later consumed.
In particular, the scope of the studies reviewed in this survey is bounded to {\em representation bias in structured (tabular) and unstructured (image, graph, text, speech) data}. 

\subsection{Related Surveys and Tutorials}
To the best of our knowledge, this is the first survey that specifically focuses on identifying and resolving representation bias in a variety of structured and unstructured types of data from a data-centric standpoint. However, we would like to highlight the existing surveys and tutorials on the general topics of bias and fairness and task-specific, data-specific, or bias-specific approaches to debiasing and promoting fairness while pointing out how their scope differentiates from our work. 

Balayn et al. \cite{balayn2021managing} is perhaps the closest study to our work in terms of scope, focusing on data-centric approaches to resolving the bias issues at the root cause, i.e., data. However, their scope is much broader in terms of the covered domains and focuses on identifying current research gaps in data management territory for tackling bias. Besides, they do not differentiate between different types of bias. This has led to interchangeably using bias and unfairness terms, while our work solely focuses on representation bias.
Moreover, \cite{balayn2021managing} mostly goes as far as introducing the works at a high level, while in our work we 
provide taxonomies and discuss technical details, with running examples where applicable.
Overall, the two surveys are in different abstract levels and have different purposes and contributions.

Stoyanovich et al. \cite{stoyanovich2020responsible} review fairness-related literature in data management pipeline in the context of automated decision systems' lifecycle. The article focuses on pre-existing, technical, and emergent bias types and how they are introduced to the data in different stages of the data management pipeline. Our survey takes a different approach by focusing on the identification and resolution of representation bias for different types of data, and focuses on the variety of data-centric techniques for these issues. Therefore, the two works have different scopes. 
For the same reasons, our work is distinguished from Catania et al. \cite{catania2022fairness} as it has a similar scope to \cite{stoyanovich2020responsible} and follows a very similar outline.

Abiteboul et al. \cite{abiteboul2019transparency} discuss a few regulatory frameworks, such as the European union's GDPR, and how the data management community can address challenges such as neutrality, fairness, data protection and transparency highlighted by these regulations.

Jagadish et al. \cite{jagadish2014big} is based on the famous white paper written by prominent researchers on the big data lifecycle and the challenges that are faced in big data analysis. However, they do not cover the responsible dimension of data analysis in their scope.

Firmani et al. \cite{firmani2019ethical} is a short paper that introduces an ethics cluster and reiterates the challenges in the information extraction pipeline associated with data quality. Our work falls into the diversity and fairness aspects of the proposed cluster. 

Mehrabi et al. \cite{mehrabi2021survey} is a comprehensive survey that classifies different kinds of bias and reviews the body of literature on machine learning fairness. We would like to highlight that \cite{mehrabi2021survey} is a complementary survey to our work that covers the general topics of bias and fairness (not necessarily data-centric) in breadth and depth. The intersection of \cite{mehrabi2021survey} and our work are only the preprocessing techniques to promote fairness as they do not consider bias measurement methods such as coverage. 

Similarly, Orphanou et al. \cite{orphanou2021mitigating} reviews the body of works on bias detection, fairness promotion, and explainability in algorithmic systems from different research communities. Overall, they have a broader scope than our work and for the same reasons as before, the intersections are only the preprocessing techniques to promote fairness. 

Finally, surveys and tutorials such as \cite{choudhary2022survey,fabbrizzi2021survey}, focus on identifying and mitigating bias within a specific research community and/or a specific type of bias and/or specific tasks.

\section{An Overview of Representation Bias}\label{sec:overview}

With the abundance of data collected from a wide range of contexts, we are transitioning from decision-making based on intuition and anecdotal observations to  decision-making based on the data. Data-driven decision-making has great potential, and success stories abound.  But there are also failures, usually because the larger volume can make it easier to hide many problems.
It is said that every decision is only as good as the data used to make it \cite{barocas2016big}.
One of the most important,  aspects of data quality is being representative of all the possible subgroups influenced by that decision \cite{firmani2019ethical}. 
This representativeness originates from how the data has been collected. With a prospective data collection approach, such as through a survey or a scientific experiment, data scientists may be able to specify requirements like representation in data. However, more often than not, data, now known as found data, is collected independently in a process that data scientists have limited or no control over. Besides, it is important to note that while data must follow the actual production distribution, this is not sufficient for the development of representative data. The data must include enough examples from "less popular regions" of data space if these regions are to be handled well by the system. 

In today's data-driven world, Automated Decision Systems (ADS) are widely used in society, ranging from fire prevention by predicting high-risk buildings to recruiting automation by screening for competitive candidates. However, historical data used for decision-making might not be objective; it could inherit historical biases in the algorithm design. For the responsible development of ADS, it is essential to analyze the representation to avoid the potential risks of injustice. For example, an attempt of the Boston government~\cite{boston} using a system to assign students to schools near their residential areas was found problematic as it ignored the fact that top schools are typically less common in underprivileged districts. For systems that rely on machine learning algorithms, without a careful inspection of the training data quality, under-representation of minority groups may cause discrimination in the prediction results~\cite{chen2018my, asudeh2019assessing, lin2020identifying,firmani2019ethical}. For example, StyleGAN~\cite{karras2019style}, one famous algorithm for auto-generating eerily realistic human faces, is also producing white faces more frequently than faces of people of color. The problem appears inherited from the training data sets, which default to white features. As a result, recent research has started to explore the relationship between machine learning bias and the inadequate sample sizes~\cite{chen2018my, ntoutsi2020bias}. Representation bias is also a crucial problem in critical domains, such as health care. First of all, there are group-specific patterns in the healthcare data. For example, many diseases are correlated with demographic factors like race, gender, etc. Ashkenazi Jewish women are known to have a higher risk of breast cancers~\cite{egan1996jewish}; the likelihood of many diseases, including obesity, hypertension, diabetes, and high total cholesterol, also varies across racial/ethnic groups~\cite{cdc}. Therefore, the medical datasets' diversity, especially demographical diversity, is vital when further using the collected data. Besides, as health data are usually sensitive, patients' willingness to share the data might vary~\cite{dash2019big}. As a result, ensuring the representativeness of the collected data is essential to avoid inaccurate or biased results in the downstream usage of the data.

\subsection{Reasons for Representation Bias}\label{sec:overview:reasons}
Bias has been studied in the statistical community for a long time ~\cite{neyman1936contributions} but social data, increasingly used for policy decision-making and by social scientists and digital humanities scholars, presents a set of different challenges~\cite{olteanu2019social,fairmlbook,barocas2016big}.
At a high level, bias in social data means certain subpopulations in data are more heavily weighted or represented due to systematic favoritism. 
It is a deviation from expectation in data and is recognized as a subtle error that sometimes goes unnoticed, causing skewed outcomes, low accuracy levels, and analytical errors. These biases are sometimes introduced to the data due to cognitive biases~\cite{harding2004cognitive,haselton2015evolution} in human reporting or flawed data collection or preprocessing. 
We refer the reader to~\cite{olteanu2019social,hammersley1997bias} for more information about the general topic of biases in social data, the origins, and various types.

The center point of this survey, representation bias, happens for a variety of reasons with no consensus on an exact set of grounds. With that in mind, we seek the origins of representation bias in one or more of the following:

\paragraph{\textbf{Historical Bias}} Historical bias is ``the already existing bias due to the socio-technical issues in the world''~\cite{mehrabi2021survey}. An example of historical bias can be found in Google's image search results. Searching for the term ``CEO United States'', the results are dominated by images of male CEOs and show fewer female CEO images. This is because only 8.1\% of Fortune 500 CEOs are women, causing the search results to be biased towards male CEOs. This problem has previously been shown for a variety of job titles, such as `CEO' in~\cite{langston2015ceo}, and Google had alleged to have resolved it. These search results are indeed reflecting reality. However, whether the search algorithms should mirror this reality or not may depend on the application and is another issue to consider.

\paragraph{\textbf{Underlying Distribution Skew}} The underlying distribution that data is collected from may lack an equal ratio or sufficient representation for all of its subpopulations. In such cases, the underlying distribution is inherently skewed, and there are no discriminatory motives behind it. For example, according to the US Census Bureau \cite{asian2019census}, around 7\% of the US population is of Asian descent while 75\% of the population is White. Collecting a uniform sample from the US society, the Asian community is considered a minority in the outcome sample and naturally less represented. However, this is a reflection of the underlying distribution which the data has been collected from. This reflection of reality may lead to discrimination against this subpopulation in some applications.

\paragraph{\textbf{Sampling/selection/self-selection Bias}} Selection bias is introduced to the data when one fails to ensure proper randomization in selecting people, groups, or tuples of data for analysis.
\textit{Sampling bias} happens on account of a non-random sampling of a population, causing some (sub) populations to 
be less likely to be sampled.
Note that selection bias is a cause for sampling bias since having selection bias, the collected samples may not represent a random sampling of a population.
\textit{Self-selection bias}, on the other hand, happens when only a subset of a selection population chooses to participate in an experiment.
This bias occurs when the intention of the participants whether to participate in the research or not creates abnormal or undesirable conditions.
Although selection bias, sampling bias, and self-selection bias are sometimes used interchangeably, it is important to differentiate between them. 
Let us clarify this distinction using an example. 
Consider a researcher who would like to conduct a survey in Chicago, mailing ballots to selected respondents.
Now if the respondents are only selected from some regions (e.g. near downtown), hence failing to ensure a random representation of different populations in the city, this is an example of {\em selection bias}.
Suppose there is no bias in the selection of respondents.
However, only a small portion of the invited respondents decide to take the survey and mail the forms back. This can cause the {\em self-selection bias}.
To see how, let us consider the famous example where the survey question is ``Do you like responding to surveys?'' with two possible options: 1) Yes, I love responding to surveys 2) No, I toss them in the trash. Now suppose only 10\% of the respondents opted to take the survey and the collected results show 99\% favored option 1.
The result is indeed invalid as the other 90\% who decided not to take the survey would likely have selected option 2!
Now, independent of how the survey was taken, if the collected samples are not random over Chicago's population, it is an instance of {\em sampling bias}.  

\subsection{Measuring Representation Bias}
In this section, we discuss the measures that have been proposed to evaluate representation bias in data.

\subsubsection{Representation Rate}
Representation rate is a metric defined in \cite{celis2020data} to identify representation bias w.r.t. the base rates.
Base rate, also known as ``prior probability'', refers to the class probability unconditioned on any observation. 
In the existing works \cite{kleinberg2016inherent,shetiya2022fairness}, an equal base rate is defined as having an equal number of objects for different subgroups in the data set. In other words, the objects in the selected set should have an equal chance of belonging to each subgroup. Consider data set $\dee$ with $n$ tuples and let $n_i$ be the number of tuples belonging to subgroup $i$. That is, for all possible subgroups $i,j$ in $\dee$, they are represented if $n_i=n_j$.

Next, we present the definition of the representation rate. 
Consider data set $\dee$ from discrete domain $\Omega:=\Omega_1\times\dots\times\Omega_d=\{0,1\}^d$ where $d$ is the number of dimensions of the dataset.
For a threshold $\tau \in (0,1]$, data set $\dee$ following the distribution $p: \Omega \rightarrow [0,1]$ is said to have representation rate of $\tau$ with respect to a sensitive attribute $\ell$ if for all $z_i, z_j \in \Omega_\ell$, we have $\frac{p[Z = z_i]}{p[Z = z_j]}\geq \tau$. 
That is, for all possible subgroups $i,j$ we have $\frac{n_i}{n_j}\geq \tau$.
The closer $\tau$ is to zero, the more biased $\dee$ is. 
Representation rate might be hard to achieve. That is because, in practice, it rarely happens that all subgroups have (almost) the same number of objects.

\subsubsection{Data Coverage}
The notion of data coverage has been studied across different settings in \cite{asudeh2019assessing,lin2020identifying,asudeh2021coverage,tae2021slice,accinelli2021impact,10.14778/3415478.3415486,accinelli2020coverage,jin2020mithracoverage} as a metric to measure representation bias. At a high level, coverage is referred to as having enough similar entries for each object in a data set. 
For a better understanding, let us go over a definition for the generalized notion of coverage. Consider a data set $\dee$ with $n$ tuples, each consisting of $d$ attributes $X=\{x_1, x_2, \cdots,x_d\}$. Attribute values may be non-ordinal categorical (e.g. {\tt race}) or continuous-valued (e.g. {\tt age}). 
Ordinal attribute values are normalized to lie in the range $[0,1]$, with values drawn from the set of rational or real numbers. 
For every tuple $t\in\dee$, $t[i]$ shows the value of $t$ on attribute $x_i\in X$.
In practice, the data scientist may be interested in studying coverage over a subset of attributes, called ``{\em attributes of interest}''. Examples of attributes of interest are {\em gender}, {\em race}, {\em salary}, etc. 
Subsequently, $X$ is assumed to be the set of attributes of interest.
The data set also contains target attributes $Y = \{ y_1,\cdots,y_{d'}\}$ that may or may not be considered for the coverage problem.

Given a query point $q\in [0,1]^d$, where $q[i]$ shows the value of $q$ with regard to $x_i\in X$, $q$ is not covered by the data set $\dee$, if there are not ``enough'' data points in $\dee$ that are representative of $q$.
In order to generalize the notion of coverage, let us define $\gee(q)$ as the group of tuples that would represent $q$. For example, suppose $X=\{${\tt gender}$\}$ and $q$ has {\tt  gender=female}. Then the set of female individuals represents $q$.
Let $\gee_\dee(q) = \gee(q)\cap \dee$. That is, $\gee_\dee(q)$ are the set of tuples in $\dee$ that represent $q$.
Using this notation, coverage of $q$ is defined as the size of $\gee_\dee(q)$. That is,
$cov(q,\dee) = | \gee_\dee(q)|$.
Given a coverage threshold value $k$, $q$ is covered if and only if $cov(q,\dee)>k$.
The {\em uncovered region} in a data set is the collection of tuples that are not covered by it. 

It is important to have a high enough coverage for all meaningful sub-populations in data regardless of the data space to make sure they are adequately represented. We would also like to emphasize the necessity of \textit{human-in-the-loop} to ignore semantically incorrect sub-populations, e.g. \{gender={\tt male}, isPregnant = \{True\}\}. Coverage thresholds are expected as an input to the problem and are supposed to be determined through statistical analyses as they are application-specific and vary by context. 
By borrowing the concept from statistics and central limit theorem, the rule of thumb suggests the number of representatives be around 30 or as \cite{sudman1976applied} suggests, for each ``minority subpopulation'' a minimum of 20 to 50 samples is necessary.

\subsubsection{Representation Rate vs. Data Coverage}
Having discussed representation rate and data coverage, let us further compare these two measures with an example. Consider a data set $\dee$ with 1000 tuples each having an attribute \textit{\{gender\}} with values \{{\tt male}, {\tt female}\}. In order to satisfy representation rate requirements, the {\tt male} and {\tt female} groups should have close counts {\em relatively to each other}.
For example, using the threshold $\tau$=0.8, 
the ratio of females-males (assuming that females are the minorities) should be at least 80\%. In other words, given that the data set size is 1K, the data set should contain at least 445 females.
On the other hand, data coverage requires a minimum count for each of the groups {\em independent from the counts on other groups}. 
So, for coverage threshold value $k$=100, each of the {\tt male} and {\tt female} groups should at least have 100 tuples to be covered. Finally, comparing the two measures, it is evident that the representation rate provides stronger guarantees of resolving issues w.r.t representation bias in downstream tasks, however, it is more restrictive and harder to achieve compared to the data coverage. In particular, when the underlying distribution is skewed (as explained in Section~\ref{sec:overview:reasons}), it is not possible to both follow the underlying distribution and fully satisfy the representation rate. 

A connection between the fairness measures and representation bias has been made
to prove fairness impossibility theorems.
In particular, Kleinberg et al.~\cite{kleinberg2016inherent} prove when there is an unequal base rate in data (i.e., representation rate is less than one), it is not possible to satisfy different fairness measures. For example, it is not possible to achieve both Equalized Odds and Predictive Parity at the same time.

\begin{figure*}[!htb] 
    \begin{subfigure}[t]{0.32\linewidth}
        	\centering
        	\includegraphics[width=\textwidth]{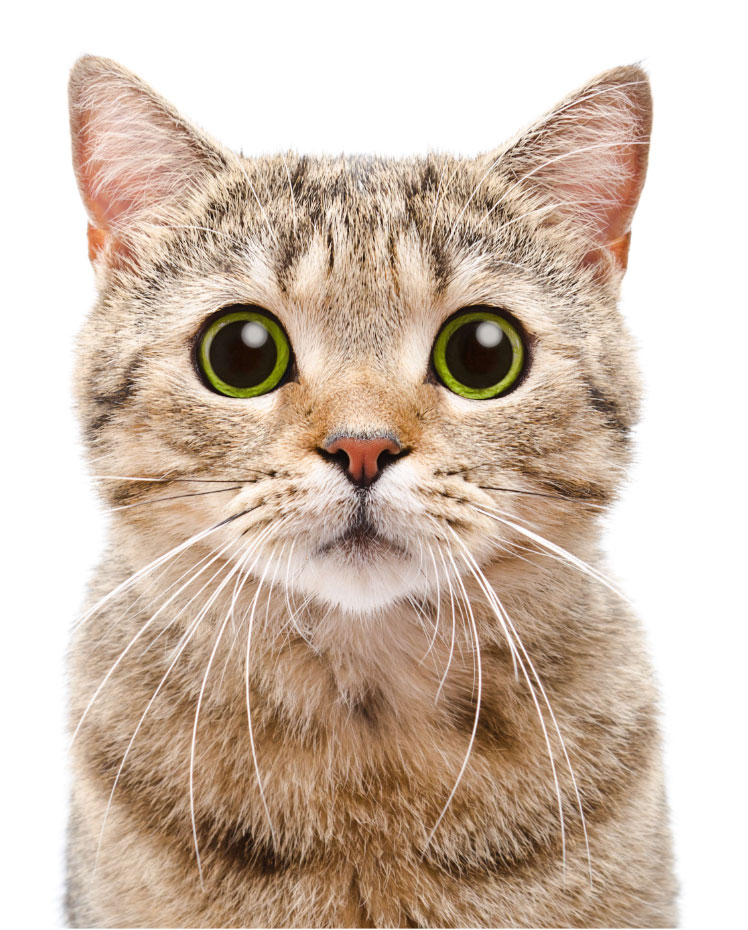} 
        	\caption{Original data set}
            \label{fig:cat-1}
    \end{subfigure}
    \hfill
    \begin{subfigure}[t]{0.32\linewidth}
        \centering
        	\includegraphics[width =\textwidth]{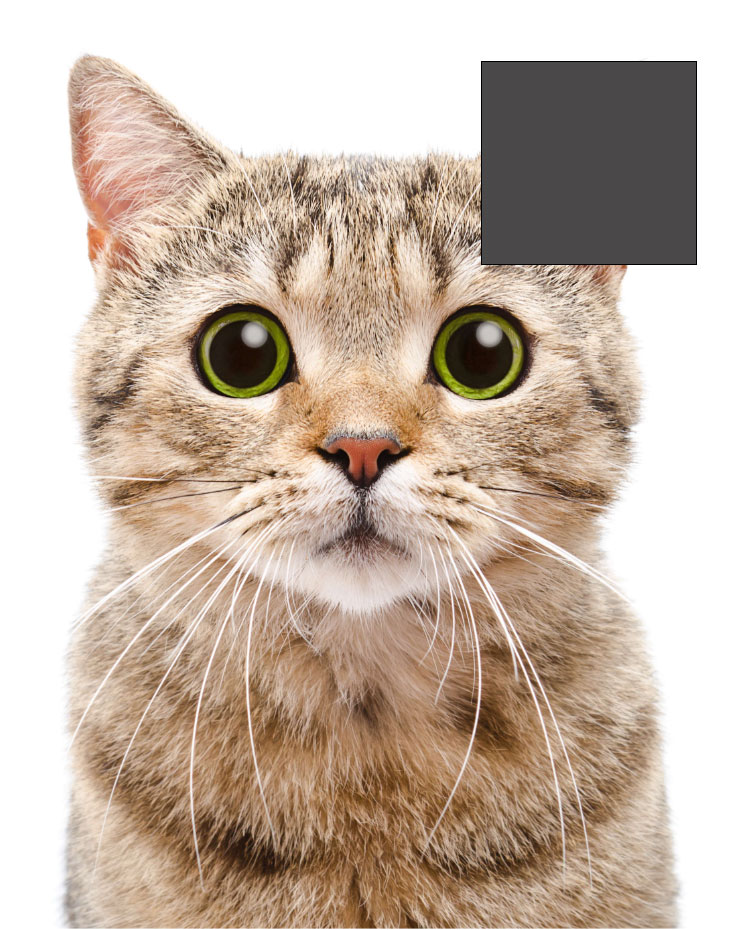}
        	\caption{Under-represented region close to decision boundary}
            \label{fig:cat-2} 
    \end{subfigure}
    \hfill
    \begin{subfigure}[t]{0.32\linewidth}
        	\centering
        	\includegraphics[width =\textwidth]{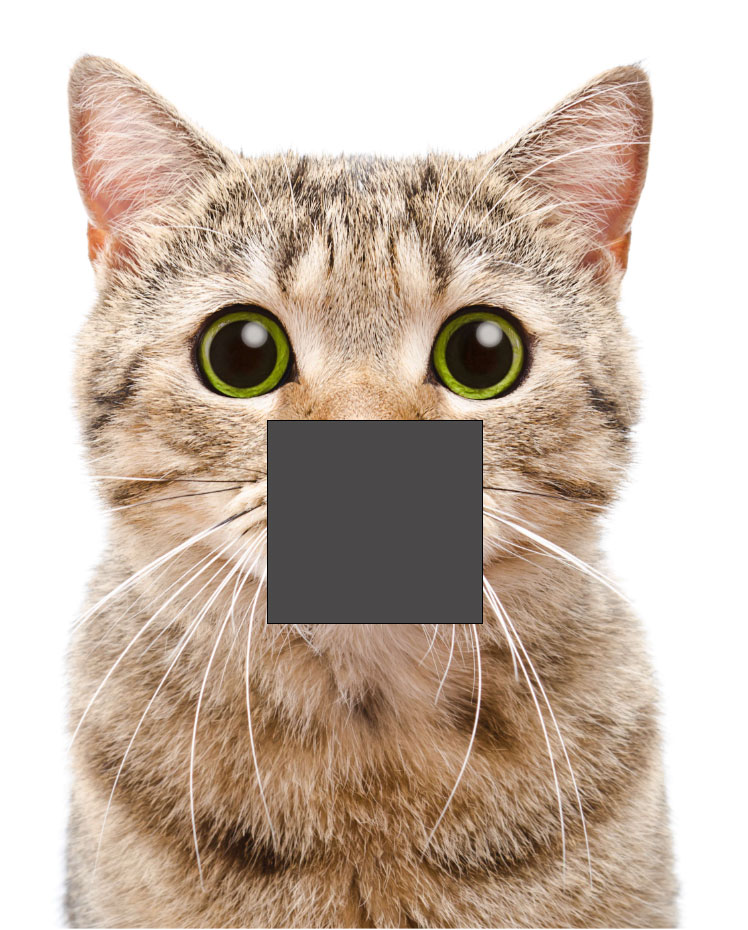}
        	\caption{Under-represented region far from the decision boundary}
            \label{fig:cat-3}
    \end{subfigure}
    \caption{Classification task: whether a query point is inside/outside the cat body. Illustration of classifier's performance for different under-represented regions.}
    \label{fig:cat}
\end{figure*} 

\subsection{Representation Bias Harms}
Before starting the discussion on representation bias identification, we would like to underscore that although representation bias is important, it does not necessarily imply poor and groundless decision-making of the system. 
For example, in a classification setting, having representation bias on continuous attributes in regions far from the {\em ground-truth} decision boundary is likely to be immaterial since those points may not contribute to refining the boundary.
Similarly, in a regression setting, in regions of the training data where the fluctuation of the target value is not much, representation bias is much less crucial than in regions with a higher fluctuation.
In general, it is safe to say that representation bias is problematic in the regions where the model behind the decision system fails to interpolate adequately based on the current data sample.

To further verify this, let us consider the following experiments, adapted from Asudeh et al. \cite{asudeh2021coverage}. First, consider a binary classification task to label a query point on the x-y plane as belonging to the body of a cat image or the background (Figure \ref{fig:cat-1}). The training data is generated by randomly sampling from the image, labeling each sample point as +1 if inside the cat's body and -1 otherwise. Next, we intentionally remove the sample points in the training data that belong to the patch highlighted in Figure \ref{fig:cat-2} to make it under-represented. Using the training data and trying multiple classification models, while the overall performance of the classifiers is high, they all fail to work for the under-represented region. In particular, while the overall false-negative rate was less than 5\%, it was as high as 54\% for the under-represented region. Relying on the training data, the models create the decision boundary by connecting the two edges of the cat's body, missing its ear. As a result, the query points that belong to the ear are misclassified as background, resulting in a high false-negative rate.

Next, we repeat the experiment, but this time, we remove the sample points belonging to the patch shown in Figure \ref{fig:cat-3}. The performance difference of the models between the overall image and the under-represented region is relatively small (around 4\%), and the model performs well for the under-represented region. Looking at the training data, the patch does not contribute to defining the decision boundary in this specific classification task and, therefore, has minimal impact on the model's performance.
\section{Representation Bias in Structured Data}\label{sec:structured}
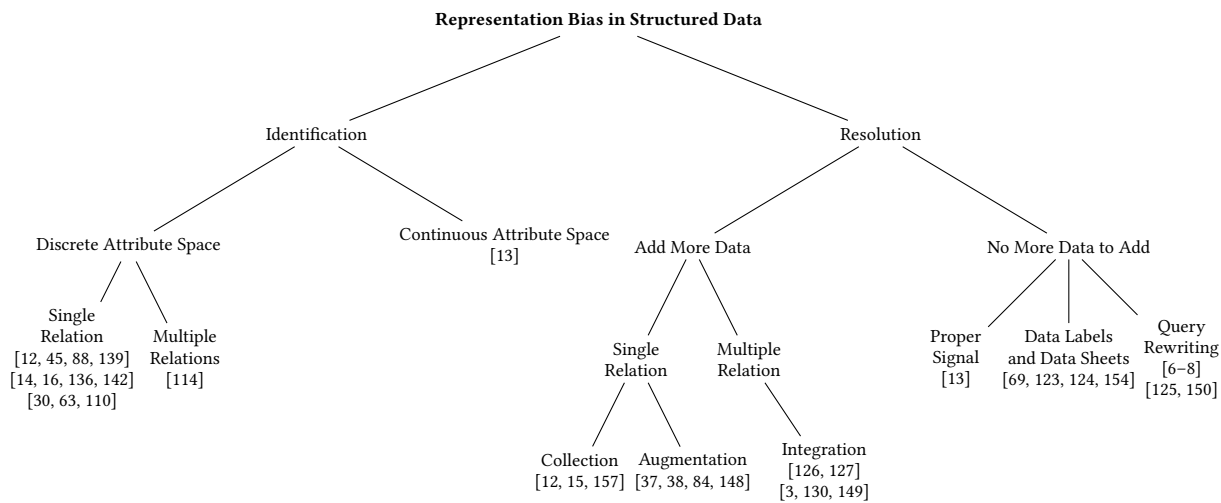
\begin{figure}[!h]
    \footnotesize
    \centering
        \begin{tikzpicture}[
            level 1/.style={sibling distance=7.5cm},
            level 2/.style={sibling distance=5cm},
            level 3/.style={sibling distance=1.5cm}]
            level 4/.style={sibling distance=1.8cm}]

            \node {\textbf{Representation Bias in Structured Data}}
                child {node {Identification}
                    child {node {Discrete Attribute Space}
                        child {node[align=center] {Single \\Relation \\\cite{asudeh2019assessing,jin2020mithracoverage,chung2019slice,pradhan2021interpretable}\\\cite{azzalini2021fair,azzalini2021short,sagadeeva2021sliceline,pastor2021looking}\\\cite{farchi2021ranking,cabrera2019fairvis,lees2019fairness}}}
                        child {node[align=center] {Multiple \\Relations \\\cite{lin2020identifying}}}
                    }
                    child {node[align=center] {Continuous Attribute Space\\\cite{asudeh2021coverage}}
                    }
                    }
                child {node {Resolution}
                    child {node {Add More Data}
                        child {node[align=center] {Single \\ Relation}
                            child {node[align=center] {Collection \\\cite{asudeh2019assessing,azzalini2021functional,tae2021slice}}}
                            child {node[align=center] {Augmentation\\\cite{sharma2020data,DBLP:journals/jair/ChawlaBHK02,iosifidis2018dealing,celis2020data}}}
                        }
                            child {node[align=center] {Multiple \\ Relation}
                            child {node[align=center, , xshift = 1cm] {Integration\\\cite{nargesian2021tailoring,nargesian2022responsible}\\\cite{abernethy2020active,shekhar2021adaptive,niss2022achieving}}}}
                    }
                    child {node {No More Data to Add}
                        child {node[align=center] {Proper \\Signal\\\cite{asudeh2021coverage}}}
                        child {node[align=center] {Data Labels \\ and Data Sheets\\\cite{gebru2018datasheets,sun2019mithralabel,Moskovitch2021PatternsCL,10.14778/3415478.3415486}}}
                        child {node[align=center] {Query \\Rewriting\\\cite{accinelli2020coverage,accinelli2021impact,accinelli2022coverage}\\\cite{shetiya2022fairness,moskovitch2022bias}}}
                    }
                };
        \end{tikzpicture}
        \caption{Classification of techniques on identifying and resolving representation bias in structured data}
\label{fig:taxonomy}
    \end{figure}

Structured data (a.k.a. tabular data) is the most common type of data available in the real world. Databases are built upon the concept of organizing data in a structured manner to facilitate tasks such as storage, querying, representation, etc. 
Representation bias in structured data has extensively been studied, and various techniques for the related problems have been proposed.
This section discusses the literature on identifying and mitigating representation bias in structured data. For each dimension, we will go through a detailed description of the research works and discuss their novelties. Figure~\ref{fig:taxonomy} depicts the taxonomy we propose for the structured data to categorize different techniques based on their objectives, capabilities, and assumptions.

\subsection{Running Example} \label{sec:running-example}

We use the Adult Income Dataset ~\cite{adult} to present running examples to better clarify the reviewed techniques. The Adult Income Data set is used to predict whether individual income exceeds \$50K/yr based on the census data.

Consider a projection of the Adult Income Dataset, shown in Figure~\ref{fig:adult}, with six attributes $\mathcal{A}=$\{\textit{gender, race, marital-status, age, hours-per-week, years-experience}\}, among which \{\textit{gender, race, marital-status}\} are non-ordinal categorical and \{\textit{age, hours-per-week, years-experience}\} are continuous-valued. The data domain for the categorical attributes are \textit{gender=}\{{\tt male}, {\tt female}\}, \textit{race=}\{{\tt White}, {\tt Black}, {\tt Asian}, {\tt Hispanic}\}, \textit{marital-status=}\{{\tt single}, {\tt married}\}.
Any attributes in $\mathcal{A}$ can be considered sensitive attributes.
The data set also contains binary ground-truth $Y = \{1,0\}$ representing whether an individual makes greater than \$50K annually or not.

\begin{figure}[!tb]
    \centering
    \begin{tabular}{|c|c|c|c|c|c|c|c|}
        \hline
         {\tt id}&{\tt gender}& {\tt race}& {\tt marital-status}& {\tt age}& {\tt hr/week}& {\tt  yrs-exp}& {\tt above-50k} \\ \hline
         1&{\tt male}& {\tt white}& {\tt single}& 21& 40& 3& 0 \\ \hline
         2&{\tt female}& {\tt white}& {\tt single}& 28& 38& 5& 0 \\ \hline
         3&{\tt male}& {\tt white}& {\tt married}& 35& 45& 10& 1 \\ \hline
         4&{\tt male}& {\tt black}& {\tt single}& 30& 40& 8& 0 \\ \hline
         \multicolumn{8}{|c|}{...} \\ \hline
    \end{tabular}
    \caption{A toy illustration of the running example (the Adult Income Dataset)}
    \label{fig:adult}
\end{figure}

\subsection{Identification of Representation Bias}
In this section, we study the works focused on identifying representation bias in structured data.
Depending on the type of the attributes of interest, we categorize the techniques into two classes based on whether they target the problem for \textit{discrete} (non-ordinal; e.g. {\tt race}, {\tt gender}) or \textit{continuous} (ordinal; e.g. {\tt age}) attributes. The attributes of interest considered for representation bias often include sensitive attributes (a.k.a. protected attributes) such as {\tt race} and {\tt gender} but are not necessarily limited to them.    

\subsubsection{Discrete Attribute Space}
Let us begin with cases where attributes for identifying representation bias are categorical.
To better observe representation bias in such cases, let us consider the following example:
\begin{example}[Representation bias in discrete attribute space]\label{example1}
Consider the running example data set (Figure~\ref{fig:adult}) described in Section \ref{sec:running-example}. Suppose the categorical attributes {\tt\small \{race, gender, marital-status\}} are used for representation bias identification.
Each conjunction of attribute-value assignment for a subset of attributes specifies a subgroup such as {\tt\small \{race=black $\wedge$ gender=female\}}.
If there are not enough tuples in the data set matching a specific subgroup,
it may not be a suitable data set on which to train a system to make a decision for that group.
\end{example}

The existing work has evaluated representation bias in discrete space using the discrete notion of coverage measure and representation rate. 
Many critical research fields have targeted the problem of identifying representation bias from different perspectives. For example, in machine learning it is important to identify under-represented subgroups in the data used to build the models as they are at a higher risk of experiencing unfairness in downstream data-driven algorithms \cite{asudeh2019assessing, jin2020mithracoverage}. Another closely related problem in machine learning is model validation by finding problematic regions in data that the model will perform poorly \cite{chung2019slice,tae2021slice,sagadeeva2021sliceline,pastor2021looking}.

Depending on whether the data is single or multiple related, in the following, we will study the techniques for identifying representation bias in discrete structured data.

\paragraph{\textbf{Single Relation}}
The majority of the existing works focus on studying representation bias for data sets that populate data in just a \textit{single} table.

We begin with \cite{asudeh2019assessing} that identifies representation bias in discrete space using the discrete notion of coverage measure.
For cases where attributes of interest are non-ordinal categorical, coverage is defined as having ``enough'' entries in the data set matching a particular pattern.
A \textit{pattern} is a string that specifies a subgroup (e.g. gender={\tt male}  $\wedge$ race={\tt white}) that matches possible values over a subset of attributes of interest.
Coverage is usually discussed for groups given by the conjunction of attribute-value assignments. A constant value is considered as the threshold for coverage, meaning that a minimum number of entries equal to the threshold value should exist from a subpopulation to be covered. 
In discrete data sets, there are multiple attributes each having multiple possible values that form a combinatorial number of possible patterns. Since patterns are the combination of some or all attributes-values, they can have multiple children and parents.
A pattern $P_1$ is the parent of pattern $P_2$, if $P_1$ can be obtained by replacing one of the deterministic elements in $P_2$ with {\tt X}. Deterministic elements in a pattern have a specified value, while non-deterministic elements are indicated by {\tt X}. As a simple example, consider a pattern defined over a single binary attribute \textit{gender} with domain \{{\tt male}, {\tt female} \}. 
Pattern $P_1$: (gender={\tt X}) is the parent to either of patterns $P_2$: (gender={\tt male}) or $P_3$: (gender={\tt female}). Equivalently patterns $P_2$ and $P_3$ are the children of $P_1$.
Depending on the size and skew in data sets, the coverage of patterns could be different and Asudeh et al. try to identify patterns that do not have sufficient coverage in an efficient way. 
If a pattern is uncovered, all of its children are also uncovered. This suggests that uncovered patterns should be identified in a way that is not dominated by more general ones, for example, if patterns $P_1$: (gender={\tt X}) and $P_2$: (gender={\tt male}) are both known to be uncovered, $P_1$ is said to dominate $P_2$ if $P_1$ is the parent of $P_2$. Uncovered patterns that do not have uncovered parents are referred to as maximal uncovered patterns (MUPs).
Therefore, the problem of identifying representation bias using the discrete notion of coverage is defined as followed: Given a data set $\dee$ defined over $d$ attributes with cardinalities $c$, as well as the coverage threshold $\tau$, try to find all MUPs.

No polynomial time algorithm can guarantee the enumeration of the entire MUPs, however, several algorithms inspired by set enumeration and the Apriori algorithm for association rule mining are proposed to efficiently address this problem.  
In this regard, Asudeh et al. introduce \textit{Pattern Graph} data structure that exploits the relationship between patterns to do less work than computing all uncovered patterns by removing the non-maximal ones. 
The parent-child relationship between the patterns is represented in a graph that can be used to find better algorithms. 
\textit{Pattern-Breaker} starts from the top of the graph where the general patterns are and moves down by breaking each pattern into more specific ones. If a pattern is uncovered, then all of its descendants are also uncovered and they can not be an MUP, even if they have a parent that is covered. Therefore, this subgraph of the pattern graph can be pruned. 
The issue with \textit{Pattern-Breaker} is that it explores the covered regions of the pattern graph and for the cases where there are a few uncovered patterns, it has to explore a large portion of the exponential-size graph. 
To tackle this, \textit{Pattern-Combiner} algorithm is proposed that performs a bottom-up traversal of the pattern graph. It uses an observation that the coverage of a node at the level of the pattern graph can be computed as the sum of the coverage values of its children. 

\begin{example}[Pattern-Combiner (Asudeh et al. 2019)] \label{example2}
Consider the subgroup race={\tt Asian} AND gender= {\tt female} in Example \ref{example1}, this data pattern is in the bottom layer of the Pattern Graph as it contains no unspecified values. It has no children and three parent data patterns: (race=X  $\wedge$  gender= {\tt female}), (race={\tt Asian}  $\wedge$  gender= X), and (race=X  $\wedge$  gender= X). If we find (race={\tt Asian}  $\wedge$  gender= {\tt female}) has enough coverage, all its parents are covered. Pattern-Combiner visits the data patterns in the Pattern Graph in a bottom-up manner, and once we find the covered pattern, we can get the coverage of its parents. 
\end{example}
The problem with \textit{Pattern-Combiner} is that it traverses over the uncovered nodes first and therefore, it will not perform well for the cases in that most of the nodes in the graph are uncovered. 
In fact, for the cases where most of the MUPs are placed in the middle of the graph, both \textit{Pattern-Breaker} and \textit{Pattern-Combiner} will not be efficient as they should traverse half of the graph. Therefore, they propose \textit{Deep-Diver}, a search algorithm based on Depth-First-Search that quickly finds the MUPs, and use them to limit the search space by pruning the nodes both dominating and dominated by the discovered MUPs.

Jin et al. \cite{jin2020mithracoverage}, design a system on top of the methods and algorithms proposed in \cite{asudeh2019assessing} to investigate representation bias over the intersection of multiple attributes using the notion of coverage.

The next work by Chung et al.~\cite{chung2019slice} proposes SliceFinder \footnote{Note that, unlike previous works, this work (as well as \cite{sagadeeva2021sliceline, pradhan2021interpretable,azzalini2021functional,azzalini2021fair,pastor2021looking,farchi2021ranking}, explained later) is model-aware. While this assumption may place these works in the scope of fairness-related literature, due to their data-centric approaches, we include them in our survey.} as a solution to address a similar problem to identifying representation bias in data. They try to determine if a model under-performs on some particular parts of data (referred to as a data \textit{slice}) since the overall model performance can fail to reflect that of smaller data slices. A slice is a conjunction of attribute-value pairs (similar to patterns in ~\cite{asudeh2019assessing}) and is considered problematic if the classification loss function takes very different values between the slice and the rest of the data. Enumeration of all possible slices is not practical and searching for the most under-performing slices can be deceptive since model performance over smaller slices can be noisy or they may be too small to have a considerable impact on the quality of the model. The goal is to identify the top-$k$ largest and most problematic slices for which the model does not perform well. 
Finding the most problematic slices requires a balance between the significance of the difference in loss and the magnitude of the slice. To do so, the disparity between the loss of a slice and its counterpart is calculated using a loss function like \textit{logarithmic loss} such that the difference is always non-negative (slice has a higher loss than its counterpart). To determine if the difference is significant, Chung et al. suggest treating each slice as a hypothesis and performing two tests to determine 1) if the loss disparity is \textit{statistically significant} (not observed by chance) and 2) whether the \textit{effect size} of the disparity is large enough (how problematic the slice is). Therefore, they find a handful of the largest problematic slices, by taking all problematic slices with an effect size larger than a threshold and ranking them by size (number of entries). 
In order to search for problematic slices, Chung et al. propose three algorithms including a baseline. 
First, they propose the \textit{Decision Tree Training} method in which, they train a decision tree to partition examples into slices defined by the tree. To find the $k$-problematic slices, they perform a Breadth-First-Search on the decision tree in which slices in each level are sorted based on an increasing number of literals, decreasing slice size, and decreasing effect size and filtered whether they are statistically significant and have large enough effect-size. The advantage of using the decision tree approach is its natural interpretability and the fact that it needs to be expanded a few levels to find the top-$k$ problematic slices. Conversely, a decision tree is optimized for classification results and may not find all problematic slices. Besides, in cases of overlapping data slices, the decision tree will find at most one of them. 
To overcome the aforementioned problems, they propose the \textit{Lattice Searching} algorithm, in which slices form a lattice and problematic slices can overlap. Lattice searching follows the same procedure as the decision tree training algorithm to search for the problematic slices. Lattice search can be more expensive than the decision tree training approach and cannot address the scalability issue of searching over the exponential size of data slices therefore, they suggest employing parallelization and sampling techniques. To better clarify how the lattice search algorithm works, let us look into an example:
\begin{example}[Lattice Search (Chung et al. 2019)]
Consider data set $\dee$ described in section \ref{sec:running-example}. For simplicity, suppose that we are interested in top-2 largest slices only w.r.t. to gender and marital-status attributes, and the effect size threshold is $T$. Initially, priority queue $Q$ includes the entire data as a slice. This slice does not have the required effect size and thus is expanded into slices gender={\tt male}, gender={\tt female}, marital-status={\tt single}, and marital-status={\tt married} that are inserted in the queue. Next, suppose that gender={\tt female} slice has the minimum effect size $T$ and is therefore dequeued and added to the top-2 results. With none of the remaining slices having an effect size $T$, the largest remaining slice (supposedly marital-status={\tt single}) is expanded. Suppose marital-status={\tt single} $\wedge$ gender={\tt male} has the minimum effect size $T$, then it is added to the top-2 results and the algorithm stops. Note that marital-status={\tt single} $\wedge$ gender={\tt female} is already considered as it is a subset of gender={\tt female} slice.
\end{example}

Next work, SliceLine~\cite{sagadeeva2021sliceline}, expands on the idea of the previous work \cite{chung2019slice} for exact slice enumeration to find real top-$k$ problematic data slices. This is due to the fact that none of the methods introduced in \cite{chung2019slice} are able to find the real top-$k$ problematic slices and this uncertainty creates trust concerns.
Utilizing frequent itemset mining algorithms and monotonicity for effective pruning, they present a sparse linear algebra implementation of slice enumeration that is efficient in practice.
To do so, a scoring function is devised that linearizes the errors and sizes by involving the ratio of average slice error to average overall error, and deducting the ratio of overall size to slice size, while weighting these segments by the user parameter $\alpha$. Using this scoring function all slices with a score larger than zero are slices of interest and will be returned in descending order of their score. They also propose upper bounds for the scoring function based on which the search lattice can be effectively pruned.

Pradhan et al. \cite{pradhan2021interpretable} propose a related approach to \cite{chung2019slice} to identify the patterns in the data that are responsible for bias from a causal perspective. Using the same notion of pattern as \cite{asudeh2019assessing}, they use interventions to measure the effect of patterns in data that significantly promote bias. To do so, they remove a subset of data that is assumed to be the root of bias and evaluate whether a classifier built on the remaining data is less discriminatory. The bias of each pattern is evaluated with the \textit{interestingness} measure. Given a fairness metric $\mathcal{F}$, data set $\mathcal{D}$ and pattern $p$, interestingness of pattern $p$ is defined as $\frac{\mathcal{F}_\mathcal{D} - \mathcal{F}_{\mathcal{D}/p}}{Sup(p)}$ where $\mathcal{F}_\mathcal{D}$ is the bias of a classifier trained on $\mathcal{D}$, $\mathcal{F}_{\mathcal{D}/p}$ is the bias of the classifier trained on intervened $\mathcal{D}$ (by removing $p$ from $\mathcal{D}$), and $Sup(p)$ is the fraction of data points that satisfy pattern $p$. In order to find the top-$k$ patterns causing the most bias, Pradhan et al. utilize a similar bottom-up approach to the lattice-based search that we saw in \cite{chung2019slice,asudeh2019assessing}.

Azzalini et al. \cite{azzalini2021fair,azzalini2021short} propose yet another related approach to detect representation bias in the data based on conditional functional dependencies (CFDs). CFDs are conditional dependencies that apply to only a subset of tuples specified with a condition. They use techniques proposed in \cite{caruccio2015relaxed} to explore the CFDs and filter out all of the CFDs that do not have at least one sensitive attribute and target variable or some of the present attributes are not assigned a constant value. Among the remaining CFDs, they calculate the difference in confidence without and with the sensitive attribute on the left-hand side. The confidence value indicates how often the CFD has been true. A positive confidence difference is indicative of bias toward the sensitive group on the left-hand side of the CFD. Finally, they rank the CFDs w.r.t. multiple criteria of interest such as support-based (number of tuples affected by the bias in CFD), difference-based (largest impact of the protected attribute on the right-hand-side), and mean-based (balance between the two prior criteria).

Pastor et al. \cite{pastor2021looking} propose the notion of \textit{divergence} to estimate different classification behavior in subgroups compared to the overall data set. Divergence measures the difference in statistics such as false-positive rate and false-negative rate between a subgroup and the entire data set. However similar to \cite{chung2019slice}, to recognize the problematic subgroups, they only consider the most frequent patterns with a size larger than a threshold and discard smaller subgroups. Once subgroups with high divergence are recognized, they check whether they are statistically significant or not due to fluctuations caused by the finite size of the data set. Next using the notion of Shapley value \cite{shapley1952project}, they investigate which attributes in each problematic subgroup are contributing the most to the local and global divergence. In this work, Shapley values measure the contribution of each attribute value to the subgroup divergence. 
\textit{DivExplorer} algorithm extracts frequent subsets of attribute values and estimates their divergence.
It begins by accepting a data set $\mathcal{D}$ including the ground-truth values, the prediction results from a model, and a support threshold value. Next, it examines each data point in $\mathcal{D}$ to be a false-positive, false-negative, or otherwise, and the results are mapped into a one-hot-encoding representation. Next, depending on the frequent pattern mining (FPM) algorithm of choice (using off-the-shelf techniques), for each step $i$ in FPM, itemsets with the minimum required support are extracted. Next, the cardinality of each itemset w.r.t. to the outcome function (false-positive, false-negative, or otherwise) is calculated. If the support of the itemset (sum of the cardinalities divided by the size of $\mathcal{D}$) is more than the specified support, threshold, the itemset is added to the list of frequents. Once all of the frequent itemsets are determined, the outcome rate of interest (false-positive rate, false-negative rate, Accuracy, etc.) is estimated for all frequent itemsets and the divergence of all frequent itemsets as the difference outcome rate for the itemset $I$ and the entire data set $\mathcal{D}$ is computed and returned. 

In another related work, Farchi et al. \cite{farchi2021ranking} propose \textit{Shapley Slice Ranking Mechanism with focus on Error concentration (SSR-E)} as an approach to rank data slices by the order of being problematic. However, they assume the slices are given as an input and they use the notion of Shapley value to rank the slices. They model the slices as players in a cooperative game and capture the importance of error concentration and statistical significance of the slices by defining various characteristic foundations. 
SSR-E accepts a model and a data set $\mathcal{D}$ with $n$ slice as input. For each slice, the algorithm calculates the set of data points that are misclassified by the input model. The Shapley value of each slice is calculated as the independent sum of the originality of its data points. The originality of each misclassified data point is proportional to the number of slices to which it belongs. Finally, the slices are returned in a non-increasing order w.r.t. their Shapley values. 

Cabrera et al. \cite{cabrera2019fairvis} propose a system called FairVis that employs a different approach to identify underrepresented subgroups in the combinatorially large space. They perform clustering on the training data set to find statistically similar subgroups and then use an entropy technique to find important features that are more dominant in that subgroup. When a feature's entropy is too close to zero, it means that it is concentrated in one value, which makes the feature more dominant in that subgroup. Next, they calculate a fairness score on the clusters and present the subgroups to the user sorted by the score. Once a problematic subgroup has been identified, users can compare them with similar subgroups to discover which value differences impact performance or to form more general subgroups with fewer features. The similarity between a pair of subgroups is calculated by summing the Jensen-Shannon divergence between all features.

Finally, in \cite{lees2019fairness}, Lees et al. suggest exploring each subpopulation's sample complexity bounds for learning an approximately fair model with a high probability. Sample complexity provides a lower limit on the count of training samples that are necessary from the subpopulations to learn a fair model.
They demonstrate that a classifier can be representative of all subgroups if adequate population samples exist and the model dimensionality is aligned with subgroup population distributions.
In case the sampling bias of the subpopulations is not met, human interventions in the data collection process by correcting representation bias (for example, collecting more data for under-represented subpopulations) are recommended.

\paragraph{\textbf{Multiple Relations}}
In the real world, data is more commonly stored and integrated into databases with \textit{multiple} tables. In order to analyze the representation bias, a combinatorial number of attribute-value combinations from different tables needs to be explored.
In this process, the data to be analyzed is obtained through complex operations, e.g., table joins and predicate combinations, in databases with multiple relations. Due to the sheer data volume, determining adequate coverage can require a prohibitively long execution time. 
In~\cite{lin2020identifying}, Lin, et al. focus on the threshold-defined coverage identification in the multiple table scenario. Following the definition of the data pattern and MUP in the single table scenario, the coverage of a pattern $P$ in a database with multiple relations is defined as the number of records satisfying $P$ in the equal join result over all the tables. 
The coverage analysis for multiple relations has two main challenges: (1) For a given data pattern $P$, to determine its coverage in the database, we need to execute a conjunctive COUNT query with table joins. It would be hard for the users to enumerate the queries for all data patterns and the execution time for a combinatorial number of such queries is prohibitive. Query optimization for the set of conjunctive COUNT queries to determine MUPs is needed for coverage analysis. 
(2) In the lattice space of the pattern graph, we need to design search algorithms to identify the set of MUPs with the minimum number of COUNT executions. The authors design a highly parallel index scheme to handle joins and cross-table predicate combinations to efficiently compute the number of records for each given group. As discussed in~\cite{asudeh2019assessing}, the MUP identification problem is an NP-hard problem. To traverse the combinatorially large search space of the pattern graph, ~\cite{lin2020identifying} designs a priority-based search algorithm that could minimize the number of computations to assess the count for a given group. The priority-based algorithm keeps searching the nodes with higher pruning efficiency. When a node is dominated by MUPs or dominates a covered pattern, it prunes this branch based on the coverage monotonicity property. The priority of the nodes is computed by a heuristic priority scoring function:
$$priority = \omega_{p} \times n_p + \omega_{c} \times n_c$$
where $n_p$ and $n_c$ are the numbers of parent nodes and child nodes for each data pattern, $\omega_{p}$ and $\omega_{c}$ are the weights for parents and children. With a higher weight for child nodes, the priority algorithm would be close to top-down BFS, while with a higher weight for parent nodes, the algorithm is more likely to traverse deep to the lower layers.

Besides, as the number of patterns does not need the exact counts for the patterns, we only need to determine whether the database contains more records than the given threshold or not. Therefore, this paper also provides a sampling-based approximate algorithm for coverage identification, which allows more efficient computation with smaller data sizes.

\begin{example}[Priority Search Algorithm (Lin et al. 2020)]
Consider the search process in Example \ref{example2}, the search of the priority-based algorithm will start from the root pattern: (race=X  $\wedge$  gender= X), where X represents unspecified values. Suppose it is covered and we need to explore its children to find the set of MUPs. Next, we evaluate all its children, suppose among its children, the pattern (race=X  $\wedge$  gender= {\tt female}) has more descendants than the pattern (race={\tt Asian}  $\wedge$  gender= X) (because of the different cardinalities of race and gender.). The priority-based algorithm will first compute the coverage of the pattern (race=X  $\wedge$  gender= {\tt female}), as once we determine its coverage, we can prune more patterns in the search process.
\end{example}

\subsubsection{Continuous Attribute Space}
Data in the real world often consists of a combination of continuous and discrete values.
To better understand representation bias in continuous data sets, let us look further into our running example:

\begin{example} [Representation bias in continuous attribute space]
Consider a model trained on data set $\dee$ described in section \ref{sec:running-example}.
While the model can discriminate w.r.t. categorical attributes like \textit{sex} and \textit{race}, it may also discriminate based on continuous-valued attributes such as \textit{age} (e.g., because most tech workers and job applicants are young). If there are not enough entries for different \textit{age} ranges (e.g. \textit{age>40}) in a data set, it may not be trained with enough data to make a decision for those ranges.
\end{example}

Regarding the example above, simple solutions like binning age into "young" and "old" can transform the continuous space into discrete. However, they may lead to coarse groupings that are sensitive to the thresholds chosen. It may be inappropriate to treat a 35-yo as young but a 36-yo as old. 

Techniques in this category assume data with continuous-valued attributes and propose solutions for identifying representation bias in such data sets.

Following a similar definition of coverage discussed earlier in \cite{asudeh2019assessing}, Asudeh et al. \cite{asudeh2021coverage} extend the notion of coverage to continuous space for identifying representation bias.
 The problem of identifying representation bias using the continuous notion of coverage is defined as follows: Given data set $\dee$ with $n$ tuples over $d$ attributes, and vicinity radius $\rho$ and coverage threshold $k$, identify the uncovered region.
A query point in continuous data space is covered if there are enough (at least $k$) data points in its $\rho$-vicinity neighborhood. $\rho$-vicinity neighborhood is the circle centered at the query point with radius $\rho$. The uncovered region is demarcated by the collection of all the uncovered query points in the space.

Depending on the number of attributes in a data set, they propose two algorithms for identifying uncovered regions in data. 
First algorithm known as \textit{Uncovered-2D} studies coverage over two-dimensional data sets where $X=\{x_1,x_2\}$. In order to find the number of circles that a query point falls into and consequently discover the uncovered region, \textit{Uncovered-2D} makes a connection to $k$-th order Voronoi diagrams.
Consider a data set $\mathcal{D}$ and its corresponding $k$-th order Voronoi diagram. For every tuple $t\in \mathcal{D}$, let $\circ_t$ be the $d$-dimensional sphere ($d$-sphere) with radius $\rho$ centered at $t$.
Consider a $k$-voronoi cell $\mathcal{V}(S)$ in the $k$-th order Voronoi diagram $V_k(\mathcal{D})$.
Any point $q$ inside the intersections of the $d$-spheres of tuples in $S$, i.e. $q\in \underset{\forall t\in S}{\cap ~\circ_t}$, is covered, while all other points in the region are uncovered.
 The algorithm starts by constructing the $k$-th order Voronoi diagram of the data set and then for each Voronoi cell $\mathcal{V}(S)$ in the diagram, it computes the intersection of the circles of the tuples in $S$ and marks the portion of $\mathcal{V}(S)$ that falls outside it as uncovered.
After identifying the uncovered region, a 2D map of $\{x_1,x_2\}$ value combinations is used to report the region to the user.

\begin{figure*}[!tb]
    \begin{minipage}[t]{0.40\linewidth}
        \centering
        \includegraphics[width=\textwidth]{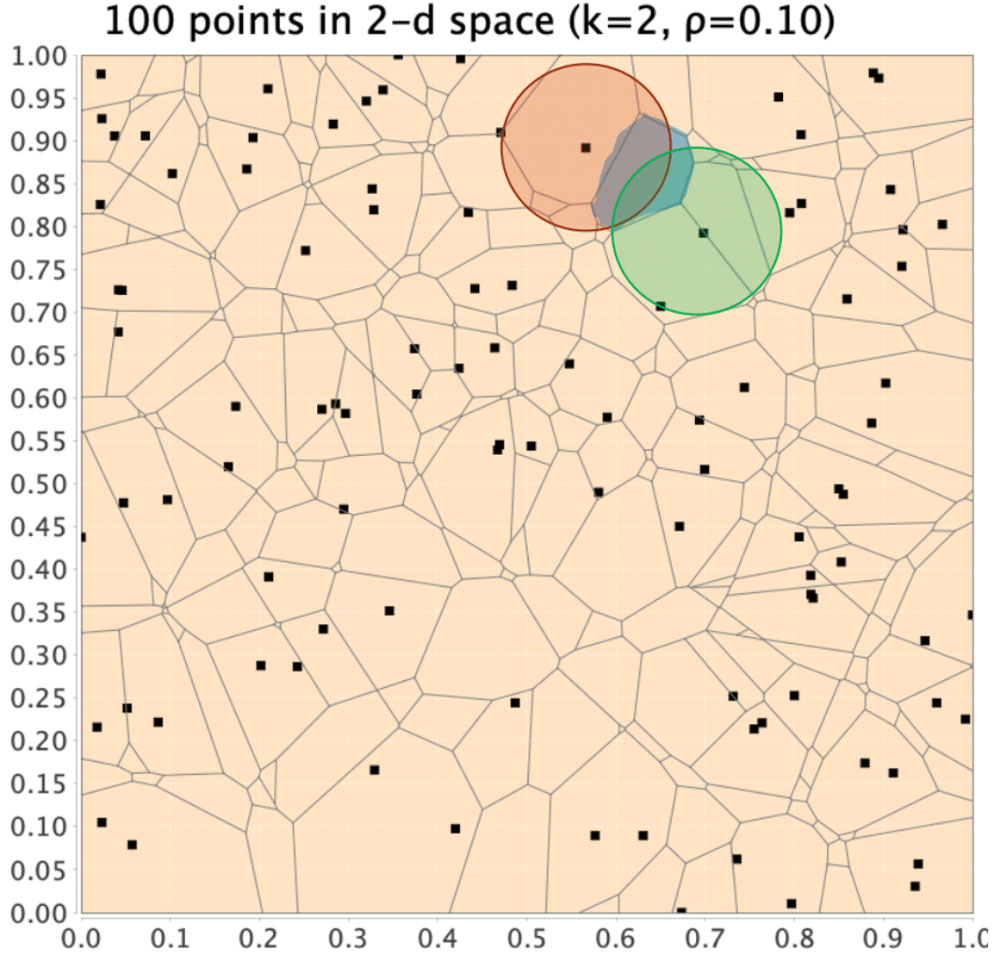}
        \caption{identifying the covered region in the gray Voronoi cell.}
        \label{fig:cvrg_2_1}
    \end{minipage}
    \hfill
    \begin{minipage}[t]{0.40\linewidth}
        \centering
        \includegraphics[width=\textwidth]{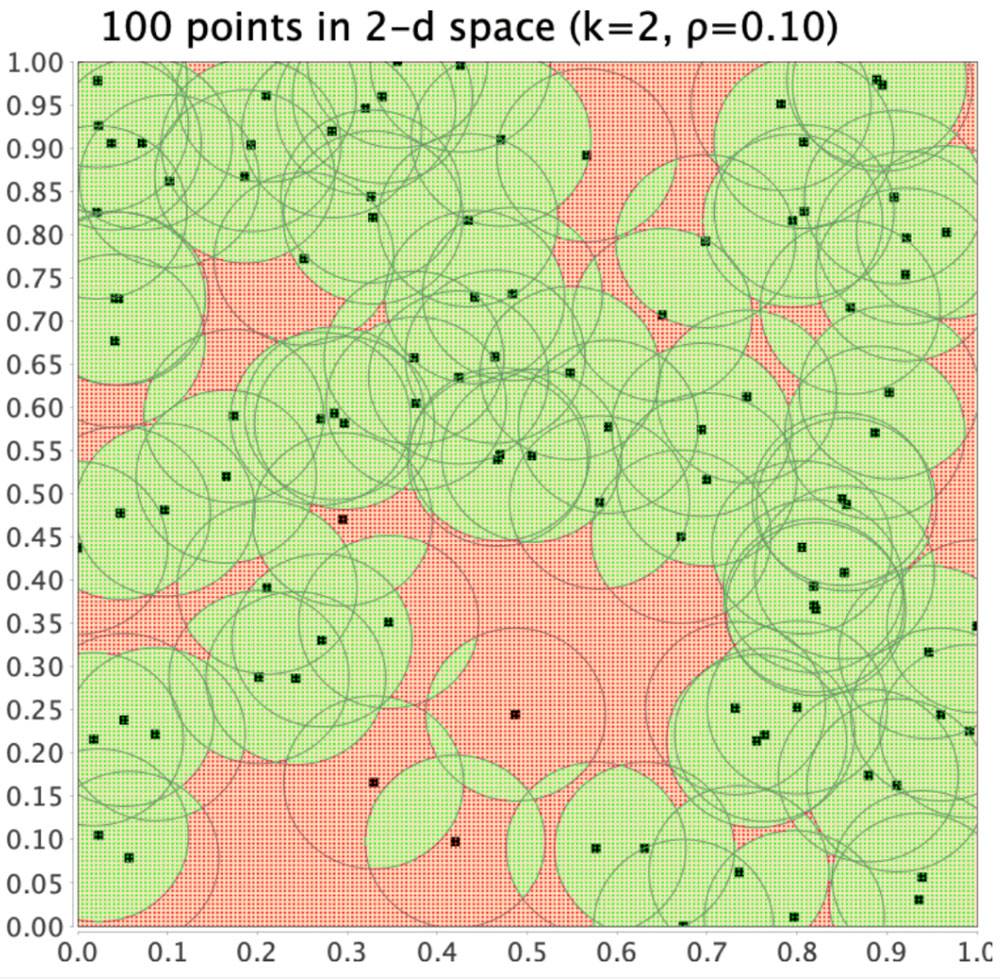}
        \caption{Covered region in data set $\dee$ marked in green. The covered region is the union of all the covered points in each Voronoi cell.}
        \label{fig:cvrg_2_2}
    \end{minipage}
\end{figure*}

Let us look into how \textit{Uncovered-2D} performs on our running example:
\begin{example}[Uncovered-2D (Asudeh et al. 2021)]
Consider data set $\dee$ described in section \ref{sec:running-example}. Suppose that we are interested in identifying the uncovered region w.r.t. to hours-per-week and years-experience attributes. Suppose that a query point is covered if it has two data points in its 0.1 radii. As illustrated in Figure \ref{fig:cvrg_2_1}, the algorithm generates the 2nd order Voronoi diagram for $\dee$, and for each Voronoi cell, the intersection of the two closest circles with radius 0.1 is considered to be the covered region. Figure \ref{fig:cvrg_2_2} shows the covered and uncovered regions in $\dee$. 
\end{example}

The algorithm for the 2D case can be extended to the general case by relaxing the assumption on the number of attributes to discover the exact uncovered region, however, due to the curse of dimensionality, the search size space explodes as the number of dimensions increases and as a result, the algorithm will not be practical. Therefore, they propose a randomized approximation algorithm based on the geometric notion of \enet \cite{haussler1986epsilon}.
In short, \enet approximates a set using a collection of simpler subsets. Let $\mathcal{X}$ be a set and $\mathcal{R}$ be a set of subsets of $\mathcal{X}$. A set $\mathcal{N}\subset \mathcal{X}$ is an \enet for $\mathcal{X}$ if for any range $r\in\mathcal{R}$, if  $|r\cap \chi|>\eps|\chi|$, then $r$ contains at least one point of $N$.
The idea is to take random samples from the space (every sample is a potential query point) and check whether each point is covered or not and label them as $+1$ if uncovered and $-1$ otherwise. If we have enough samples in this collection, an \enet is formed using which the uncovered region can be learned. The problem with \textit{Uncovered-MD} is that theoretically speaking, in adversarial cases, the number of samples may be exponentially large to the number of dimensions. However, in practice, the adversarial case is unlikely to happen since the boundary complexity depends on the number of arcs constructing it which can be significantly less than the theoretical upper bound provided for the number of samples.  

\subsection{Resolving Representation Bias}
After identifying representation bias in data, the next step is presenting a remedy for it. The first approach to tackling this problem is adding more data while hoping to address the under-representation issues. However, with limited control over the data collection processes, it could be difficult and expensive for the data scientist to collect more data from the data sources. 
When adding more data is not feasible, the current research suggests preventive solutions such as informing the user about the representation bias issue or rewriting queries to meet the representation constraints. 
With that being said, we would like to emphasize the necessity of human-in-the-loop in the resolution process. It is vital to notice that not all the under-represented regions in the data are meaningful, and some may even be invalid. Therefore, a domain expert must evaluate and semantically validate the identified groups/regions.

Generally speaking, resolution techniques that operate by adding more samples to the data set (e.g. data collection, data integration, etc.) require additional sources of data available that can be employed to resolve representation bias. Such techniques are effective when representation bias is due to the reasons such as sampling and selection bias.
On the other hand,
when reasons such as historical bias cause representation bias, it is costly, if not unlikely, to find additional sources to collect enough data from minorities.
In such cases, preventive techniques (e.g. generating warning signals, nutritional labels, query rewriting, etc.) are effective to help the users make informed decisions.

In the following, we will introduce state-of-the-art techniques for resolving representation bias in structured data. 

\subsubsection{Adding More Data}
Enriching the data set with more data is the best way to address the under-representation issues. However, adding more data is not free.  
In particular, when the representation bias is due to the underlying distribution skew (see Section \ref{sec:overview:reasons}), collecting more data from the under-represented groups
may violate the i.i.d sample requirement, as the data may no longer follow the underlying distribution.
Furthermore, there are not always opportunities for adding more data through data collection or integration. In these cases, the existing research has acquired techniques like data augmentation to potentially improve whatever data is available and address the lack of representation issues. 

\paragraph{\textbf{Data Collection}}
Data collection is usually costly. If the data are obtained from some third party, there may be a direct monetary payment. If the data are directly collected, there may be a data collection cost. In all cases, there is a cost to cleaning, storing, and indexing the data. To minimize these costs, as little additional data as possible should be acquired to meet the representation constraints.

In this regard, Asudeh et al.~\cite{asudeh2019assessing} suggest identifying the smallest number of additional data points needed to hit all the large uncovered spaces. Given the combinatorial number of patterns, it is not feasible to cover all of the patterns in practice. To do so, they determine the patterns for the minimum number of items that must be added to the data set to reach a desired maximum covered level or to cover all patterns with at least a specified minimum value count. This problem translates to a hitting set instance which can be viewed as a bipartite graph with the value combinations on the left side and the uncovered patterns on the right. There is an edge between a combination and a pattern if the combination matches the pattern. The objective is to select the minimum number of nodes on the left side that hit all the patterns on the right. The hitting set problem is NP-complete and the greedy approach to select the value combination that hits the maximum number of un-hit patterns guarantees a logarithmic approximation ratio for it. 
\begin{example}[Coverage Enhancement (Asudeh et al. 2019)]
Consider the example in Section 4.1, consider attributes \{race, marital-status, gender\}, suppose the set of MUPs contains two patterns: $P_1$: (race=X  $\wedge$ marital-status ={\tt Single}  $\wedge$  gender= {\tt female}) and $P_2$: (race={\tt Asian}  $\wedge$ marital-status =X  $\wedge$  gender= {\tt female}). A run of the greedy algorithm picks a pattern (race={\tt Asian}  $\wedge$  marital-status ={\tt Single}  $\wedge$  gender= {\tt female}), and this pattern hits both $P_1$ and $P_2$ in MUPs, therefore, the coverage enhancement process finishes.
\end{example}

Azzalini et al. \cite{azzalini2021functional} propose an approach to mitigate the representation bias in the data by adding tuples to the CFDs identified using the techniques in \cite{azzalini2021fair,azzalini2021short}. 
There are two ways to add tuples with regard to a CFD. The first option is by adding tuples to the opposite target variable of the identified CFD. As an example if \textit{ (gender = {\tt female}, marital-status = {\tt single}) $\longrightarrow$ Income = `$\leq$50K'} is the identified CFD then tuples should be added to \textit{ (gender = {\tt female}, marital-status = {\tt single}) $\longrightarrow$ income = `$>$50K'}. The second alternative is adding to advantaged group\textit{ (gender = {\tt male}, marital-status = {\tt single}) $\longrightarrow$ income = `$\leq$50K'}, however, this method could cause potential issues such as increased discrimination.
The proposed algorithm to optimally add tuples to the data set is an improved version of \textit{Greedy Hit-Count} algorithm \cite{asudeh2019assessing}. For each CFD with the opposite target variable, a vector of size $d$ ($d$ being the number of attributes in the data set) is created, and the values of the vector are filled according to the values in the corresponding CFD or $X$ if unspecified. Next, \textit{Greedy Hit-Count} algorithm accepts the discovered patterns as inputs and returns the minimum set of tuples required to repair the data set. After this step, some of the identified CFDs may still be present which can cause bias, so, in the final step, a correction algorithm removes the tuples associated with the remaining CFDs from the data set. 

In another work, Tae et al.~\cite{tae2021slice} focus on acquiring the right amount of data for data slices such that both accuracy and fairness are improved. Acquiring the same amount of data for all slices may not have the same cost-benefit and it can bias the data and affect the model's accuracy for other regions. Therefore, they propose a few data acquisition strategies (including 3 baselines) such that the models are accurate and fair for different slices. Baselines include acquiring the same amount of data for all slices, acquiring data for all slices such that in the end they all have the same amount of data (Water filling algorithm), and acquiring data in proportion to the original data distribution. None of the baselines solve the problem in an optimal way and in many cases increase the loss and unfairness of the models. This leads to the selective data acquisition problem that is defined as given a data set, a set of data slices, a model trained on the data set, a cost function for data acquisition, and a data acquisition budget, acquire examples for each slice such that the model's average loss and average unfairness over all slices are minimized while the overall cost for data collection fits the budget. 
The idea is to estimate the learning curves of slices, which reveal the cost benefits of data acquisition. The impact of data acquisition on the model's loss is significant at first but then gradually stabilizes to the point where it is not worth the effort anymore. Given the learning curves, \textit{Slice Tuner} uses the learning curves to determine how much data to acquire per slice in order to optimize the model accuracy and fairness across the slices while using a limited data acquisition budget. However, in reality, learning curves are not perfectly generated because slices may not have sufficient data for the model loss to be measured. Besides, acquiring data for one slice may affect the loss of the model on some other slices and eventually change their learning curves. So it is important to generate learning curves that are reliable enough to still benefit Slice Tuner given these issues. 
The selective data acquisition problem can be considered in two different settings: for the cases where slices are independent of each other, it is only needed to solve the optimization problem once. Since the objective for minimizing loss and unfairness is global, optimization should be done on all slices. The \textit{One-shot} algorithm updates the learning curves and solves the optimization problem to determine the amount of data that needs to be acquired for each slice. When slices are dependent Slice Tuner iteratively updates the learning curves as more data is acquired. Besides, the iterative updates make the learning curves more reliable, as they are updated whenever enough influence happens, irrespective of its direction. The \textit{Iterative algorithm} limits the change of imbalance ratio to determine the amount of data to obtain for each slice.
 Next, let us look into an example of how the Iterative algorithm works:
\begin{example}[Iterative Algorithm (Tae et al. 2021)]
Recall data set $\dee$ from section \ref{sec:running-example}. Using the aforementioned techniques from \cite{chung2019slice} slices $S_1$ of initial size 5 and $S_2$ of initial size 10 have been identified as the problematic slices. Suppose that the minimum slice size is required to be $L=10$ and the data acquisition budget is $B=55$. First, the iterative algorithm acquires 5 tuples for $S_1$ to meet the required slice size criteria which brings down the budget $B$ to 50 and updates the slice sizes for $S_1$ and $S_2$ to $[10,10]$. Next, the imbalance ratio is calculated as $\frac{10}{10}=1$. While there is still some budget left, suppose OneShot determines $[10,40]$ tuples to be acquired for $S_1$ and $S_2$. If all of this data is acquired the imbalance ratio will become $\frac{10+40}{10+10}=2.5$. Therefore, the difference between the imbalance ratio before and after data acquisition is $2.5-1=1.5$ which exceeds $T=1$ (for simplicity $T$ is a given constant). To avoid exceeding $T$, change ratio $x$ is calculated such that $\frac{10+40x}{10+10x}=2$. With $x=0.5$, the number of tuples to be acquired becomes $0.5\times[10,40]=[5,20]$. Next, the data is acquired and budget $B$, and the rest of the corresponding variables are updated and so long as there is still budget left another iteration of OneShot and the subsequent steps are executed.
\end{example}

\paragraph{\textbf{Data Augmentation.}}
Data augmentation techniques increase the size of data by adding partially altered duplicates of already existing tuples or generating new synthetic entries from existing data. Some of the existing works adopt these techniques by adding synthetic points with different values for the attribute of interest for representation. Consequently, the new data set has an equal number of elements for different values of the attribute of interest, resulting in potentially resolving the under-representation issues. 

In~\cite{sharma2020data}, Sharma et al. propose a novel data augmentation method to address the lack of representation of subgroups in a data set. For a data set with a protected attribute having a privileged and unprivileged subpopulation, they create an ideal world data set: for every data sample, a new sample is created that has the same label and features as the original sample except that it has the opposite value for the sensitive attribute compared to the original sample (e.g. if the original sample has the sensitive attribute \textit{gender=}{\tt male}, the new sample is \textit{gender=}{\tt female} and identical to the original sample w.r.t. the remaining attributes). The synthetic tuples are then sorted in order of their closeness to the original training distribution and added to the real data set to create intermediate data sets. As a result, this new data set has an equal number of entries for privileged and unprivileged sub-populations, while the label is not dependent on the protected attribute anymore, therefore potentially removing representation bias from the model built on the data set. Although, there is concern about polluting the data set with too many synthetic entries, by selectively adding the synthetic points that are closest to the original distribution in every increment. The user can see the effect of an augmentation technique that improves fairness while keeping the overall accuracy nearly constant.

Sometimes, the real-world training data could predominately be composed of majority examples with a small percentage of outliers or interesting minorities. For example, in applications like fraud detection, disease diagnoses, and the detection of oil spills, the majority of the records are negative while there is a small number of positive ``interesting'' records. Machine learning models trained on such imbalanced data sets are highly likely to have poor performance. Oversampling is one of the most commonly used methods to enhance the model performance in this case. The naive uniform oversampling algorithms simply duplicate the minorities uniformly at random and are subject to a higher risk of model over-fitting. The \textit{Synthetic Minority Oversampling Technique (SMOTE)}~\cite{DBLP:journals/jair/ChawlaBHK02} is a better alternative, which generates synthetic records of minorities based on their $k$-Nearest minority neighbors. There is a rich line of works that extend the SMOTE algorithm, for example, the SMOTE-borderline algorithms~\cite{DBLP:conf/icic/HanWM05}, which classified the minorities into noise, danger, and safe and only uses the danger minorities for data augmentation; and the extension of SMOTE for high-dimensional data~\cite{DBLP:journals/bmcbi/BlagusL13a}.

Similarly, Iosifidis et al. ~\cite{iosifidis2018dealing} suggest two techniques for resolving representation bias including an oversampling baseline by duplicating the instances from the minority subgroups to achieve balance. The idea of their main approach is to use SMOTE as an augmentation technique. They propose two approaches to creating the instances, first, producing instances based on a given attribute and populating the minority subgroup for a given attribute. Second, by generating instances based on a given attribute w.r.t. class, meaning that instances from the under-represented subgroup of a given attribute are generated to deal with the subgroup's class imbalance.

Finally, Celis et al. \cite{celis2020data} present a data preprocessing method for mitigating representation bias. The goal of this approach is to learn a distribution that resolves representation bias while remaining as close as possible to the original distribution. Learning a distribution in polynomial time to the dimension of the domain (versus domain size that can be exponential) guarantees the scalability of their method. They propose a framework based on the maximum entropy principle claiming that of all the distributions satisfying observed constraints, the distribution should be chosen that is ``maximally non-committal'' with regard to the current state of knowledge meaning that it makes the fewest assumptions about the true distribution of the data. Using this principle, probabilistic models of data are learned from samples by obtaining the distribution over the domain that minimizes the KL-divergence with regards to a ``prior'' distribution such that its expectation follows the empirical average derived from the samples.
Their approach for preprocessing data benefits from the maximum entropy framework by combining re-weighting and optimization approaches. Maximum entropy frameworks can be specified by a prior distribution and a marginal vector, providing a simple way to enforce constraints for sufficient representation.
Using a re-weighting algorithm, Celis et al. specify the prior distribution by carefully choosing weights for each tuple such that desired fairness measures are satisfied and data is debiased from representation bias. 
Let us explain the re-weighting algorithm with an example: 
\begin{example}[Re-weighting (Celis et al. 2020)]
Consider data set $\dee$ from section \ref{sec:running-example}. Suppose that 3 tuples in $\dee$ make greater than 50K per year (class positive) and 4 tuples belonging to class make less than the amount (class negative). In the positive class, the gender of 2 of the tuples are {\tt male} and 1 is {\tt female}. In the negative class, the gender of 2 of the tuples are {\tt male} and 2 are {\tt female}. The weight of each tuple $t$ is calculated as the number of tuples belonging to the class of $t$ divided by the number of tuples belonging to the same class with identical gender as $t$. Therefore the assigned weights for the tuples in $\dee$ are calculated as followed:

t= {\tt female}-positive $\rightarrow$ $w(t)=\frac{c(positive)}{c(positive, female)}=\frac{3}{1}=3$,
t= {\tt male}-positive $\rightarrow$ $w(t)=\frac{c(positive)}{c(positive, male)}=\frac{3}{2}=1.5$

t= {\tt female}-negative $\rightarrow$ $w(t)=\frac{c(negative)}{c(negative, female)}=\frac{4}{2}=2$,
t= {\tt male}-positive $\rightarrow$ $w(t)=\frac{c(negative)}{c(negative, male)}=\frac{4}{2}=2$

\end{example}
Next, a marginal vector is chosen as the weighted average vector of samples to meet the representation rate constraints. Having defined the optimization program, they solve the dual form using the Ellipsoid algorithm as it can be done in polynomial time in the dimension of data.

\paragraph{\textbf{Data Integration.}}
In data integration, data is consolidated from different sources into a single, unified view. Thus, it is a very effective solution to acquire data from different distributions such that sufficient representation is ensured for the underlying populations. However, there are sampling policy and cost-efficiency concerns that need to be examined.  

In this regard, Nargesian et al.~\cite{nargesian2021tailoring,nargesian2022responsible} suggest \textit{Data Distribution Tailoring (DT)} as resolving insufficient representation of subgroups in a data set by integrating data from multiple sources in the most cost-effective manner such that subgroups in the data set meet the count distribution specified by the user. Depending on our knowledge about data source distributions, \textit{DT} can be defined from two different perspectives, first, when the user is aware of the data source sizes and the total number of tuples belonging to each subgroup, and second, when such knowledge about the data sources do not exist. 
For the cases when the group distributions are known, the process of collecting the target data set is a sequence of iterative steps, where at every step, the algorithm chooses a data source, queries it, and if the obtained tuple contributes to one of the groups for which the count requirement is not yet fulfilled, it is kept, otherwise discarded. To do so, they first propose a \textit{Dynamic Programming (DP)} algorithm. An optimal source at each iteration minimizes the sum of its sampling cost plus the expected cost of collecting the remaining required groups, based on its sampling outcome. The dynamic programming analysis evaluates this cost recursively by considering all future sampling outcomes and selecting the optimal source in each iteration accordingly. 
\begin{example}[Dynamic Programming Algorithm (Nargesian et al. 2021)] 
Consider the data set schema from Section \ref{sec:running-example}. Suppose, to enrich the dataset, one would like to collect more samples from external sources $\mathcal{D}_1$ and $\mathcal{D}_2$.
$\mathcal{D}_1$ has 20\% {\tt female} and 80\% {\tt male}, and the sampling cost of 2. $\mathcal{D}_2$ has 40\% {\tt female} and 60\% {\tt male}, and the sampling cost of 3. For simplicity, suppose that we want to collect one tuple for each demographic group. The DP algorithm calculates the optimal cost $F(female=1,male=1)$, and decides the optimal source to query, as follows:
\begin{align*}
F(0,0)&=0 \\
F(1,0)&=min(\frac{2}{0.2},\frac{3}{0.4})=7.5 \Rightarrow \mbox{query } {\mathcal{D}_2}, \\
F(0,1)&=min(\frac{2}{0.8},\frac{3}{0.6})=2.5 \Rightarrow \mbox{query } {\mathcal{D}_1}, \\
F(1,1)&=min(2+0.2F(0,1)+0.8F(1,0)\;,\;3+0.4F(0,1)+0.6F(1,0))=8.5 \Rightarrow \mbox{query } {\mathcal{D}_1}
\end{align*}
\end{example}
The drawback to the DP algorithm is that it quickly becomes intractable for cases where the minimum count requirements for the groups are not small. However, they provide a special case for when the (sensitive) attribute of interest is binary like \textit{gender (male, female)} and the cost to query data is similar from all sources.
The authors prove that the optimal selection for this special case is to query the data source with {\em maximum probability of obtaining a sample from the minority group}.
 Similar to the previous algorithm, the process of collecting the target data is a sequence of iterations where, at every iteration, we should select a data source to query. 
At each iteration, the algorithm finds corresponding data sources for each group, and then depending
on which group is in the minority, it queries the proper data source. The algorithm stops when the count requirements of both groups are satisfied and then returns the target data set. 
Finally, as an alternative to the DP algorithm, they propose an approximation algorithm for the general case. They model the problem as $m$ instances of the ``coupon collector’s problem'', where every $j$-th instance aims to collect samples from the $j$-th group, and then using the union bound, they come up with an upper-bound on the expected cost of this algorithm. The algorithm first identifies the minority groups and then queries its corresponding data source and updates the target data accordingly. 
Let us look into a simple example:
\begin{example}[Coupon Collector's (Nargesian et al. 2021)]
Consider a case that we desire to collect 100 tuples for group $\mathcal{G}_1$ from the most cost-effective data source for $\mathcal{G}_1$ a.k.a. data source $\mathcal{D}$ that has the largest $\frac{N_1}{N.C}$ ($N_1$ is the number of tuples belonging to $\mathcal{G}_1$, $N$ is the entire number of tuples in $\mathcal{D}$ and $C$ is the sampling cost). Suppose that $N=1000$, $C=1$ and $N_1=200$, therefore, the cost to collect $\mathcal{Q}_1=100$ samples from $\mathcal{G}_1$ is bounded by $N.C.\ln{\frac{N_1}{N_1-\mathcal{Q}_1}}\simeq 693$.
\end{example}
For the cases where the group distributions are unknown, Nargesian et al. model DT as a multi-armed bandit problem. Every data source is an arm and we want to select arms in order to collect the required tuples for each group. Every arm has an unknown distribution of different groups and a query to an arm has a cost. As the bandit strategy, they adopt ``Upper Confidence Bound (UCB)'' to balance exploration and exploitation. At every iteration, for every arm, UCB computes confidence intervals for the expected reward and selects the arm with the maximum upper bound of reward to be explored next. Finally, they argue that the reward of obtaining a tuple from a group is proportional to how rare this group is across different data sources or in other words, what the expected cost one needs to pay is in order to collect a tuple from that group. 

Abernethy et al.\cite{abernethy2020adaptive,abernethy2020active} propose an adaptive sampling algorithm that adequately represents sensitive demographic groups compared to the remaining groups. In each round, the algorithm either samples from the entire population or the population that is under-represented thus far. The decision to sample from which population depends on a sampling probability value $p$ which decides whether to minimize the performance loss of the model trained on the current data ($p=1$) or minimize the fairness loss w.r.t. the under-represented group ($p=0$). With that being said, algorithm samples with a probability of $1-p$ from the under-represented group and with a probability of $p$ from the entire population. Next, the sampled point will be added to the training data, and the algorithm proceeds to the next round.  

Shekhar et al. \cite{shekhar2021adaptive} propose a similar adaptive sampling approach to \cite{abernethy2020adaptive} based on the optimism principle to actively create a data set that converges to min-max fair solutions. The optimism principle is used in the multi-armed bandit literature and tries to identify the hardest group to choose from. Given a fixed amount of budget, the algorithm dedicates more from the budget to the hardest groups (disadvantaged groups performing worst) w.r.t. a sensitive attribute and samples more from their distribution. 

While \cite{abernethy2020adaptive, shekhar2021adaptive} assumes that collecting a fair data set from existing sources is always attainable, this assumption may not always hold.
In this regard, Niss et al. \cite{niss2022achieving} propose an approach to check the feasibility of collecting a data set from a set of available sources such that the minority groups are properly represented. To do so, the adaptive sampling is reduced to the convex hull feasibility problem which is to determine whether a point falls in the convex hull of the means from a set of unknown distributions. Given a known variable $x$ and a confidence value $\epsilon>0$ and open set $x_{\epsilon}=\{y:||y-x||<\epsilon\}$, a sampling policy is feasible if there exists a $y\in x_{\epsilon}$ that lies in the convex hull of the means and otherwise infeasible.
They study the convex hull feasibility problem in Bernoulli and Multinomial settings and devise four sampling algorithms as followed: 
The \textit{uniform} algorithm at each iteration chooses from the distribution with the least samples resulting in a uniform sample size for all distributions. 
\textit{LUCB Mean} chooses from the distribution with the confidence boundary farthest from $x$ in the direction of greatest uncertainty.
The direction of greatest uncertainty is the direction away from $x$, a distribution mean is least likely to lie on.
\textit{LUCB Ratio} chooses from the distribution whose confidence region has the biggest fraction of area on the side of $x$ in the direction of greatest uncertainty. 
\textit{Thompson Sampling} commonly used in the multi-armed bandit literature, samples a mean from the posterior of each distribution, and chooses the distribution with the mean furthest from $x$ in the direction of greatest uncertainty.

\subsubsection{No More Data Available to Add}
It is not always possible to add more data to the data sets as there might be complications such as unknown underlying distribution, lack of additional data, etc. Existing work suggests alternative solutions to tackle these scenarios, such as informing the users about the deficiencies in the data set or raising warnings at query time. Furthermore, by adding proper constraints on the queries w.r.t. the attributes of interest, an effort is made to ensure the proper representation. 

\paragraph{\textbf{Generating Proper Warning Signal.}}
Generating proper signals for the trustworthiness of the analysis \cite{asudeh2021coverage} occurs when querying about a particular data point that might potentially be concerning due to belonging to an under-represented subpopulation. 
The warning signal states whether the query point is covered or not. For the 2D case, the idea is to find the Voronoi cell that the query point belongs to and check the point's distance to all the points from the data set that fall into that cell. If either of the distances is larger than the vicinity threshold, the query point is uncovered and a warning signal is generated. For the MD case, the classifier trained on the last iteration of the \textit{Uncovered-MD} algorithm is used to determine the coverage of the query point by the data set.
Finally, whether to consider the outcome and how to take action is a decision left to the model user.

\paragraph{\textbf{Data Labels and Data Sheets.}}
Annotating data sets with representation information informs the data scientist about the potential deficiencies due to representation bias when the model is being constructed. This is a signal to investigate the fitness of data for a particular task before building the models.

In ~\cite{gebru2018datasheets}, Gebru et al. propose a list of questions that data set collectors should have in mind before the procedure and respond to after the collection is done. Users can then make informed decisions about the fitness of the data set for their tasks. A number of these questions address the representativeness of the data set such as whether the data set includes all possible instances or is a sample (not necessarily random) of a larger set and if it is the latter, what is the larger set? Is the sample representative of the larger set and if so how the representativeness was verified, otherwise, why not? Does the data set identify any subpopulations such as race, gender, age group, etc., and, if so, how are these subpopulations identified, and what is their distribution like in the data set? Does the data set include attributes that can be considered sensitive like racial or ethnic origins, sexual orientations, religious beliefs, political opinions, etc?

Some research proposes using data labels to help data users choose the appropriate datasets for their tasks. Information about data coverage is important to the data set profiling.
MithraLabel~\cite{sun2019mithralabel} provides a set of visual widgets delivering information about the data set among different tasks on the representativeness of minorities, bias, correctness, coverage in terms of MUPs, outliers, and much more. 

In~\cite{10.14778/3415478.3415486, Moskovitch2021PatternsCL}, Moskovitch et al. design a ``coverage label’’ of compact size that can be used to efficiently estimate the counts for each combination of discrete attributes (pattern). They provide a trade-off between the label size and the estimation error of pattern counts. The label model is built upon an estimation function that allows the users to estimate the count of every pattern. The authors design a label for a given subset $S$ which stores the pattern count for each possible pattern over $S$ and the value count of each value appearing in the data set. The identification of the optimal labels is an NP-hard problem. The authors also present an optimized heuristic for optimal label generation.

\paragraph{\textbf{Query Rewriting.}}
Consider a data set with some interesting attributes (for example, gender, race, age) that are prone to be under-represented and a query over the data. Now suppose that some representation constraints are given w.r.t. the result of a query when executed over the data set (for example, the number of females to be greater than a given threshold), but when the query is executed over the data set, results do not satisfy the required constraints.
The idea of query rewriting is to minimally rewrite the transformation queries so that certain representation constraints are guaranteed to be satisfied in the result of the transformation.

Accinelli et al.~\cite{accinelli2020coverage} propose an approach for rewriting filter and merge operations in preprocessing pipelines into the closest operation so that the unprivileged groups are sufficiently represented. This is motivated by the fact that the under-representation of a subpopulation in an initial or intermediate data set in preprocessing pipelines may lead to the under-representation of that subpopulation in any future analyses. To do so, they provide an approach that minimally rewrites the transformation operation such that coverage constraints are ensured to be met in the transformed outcome. Many potential rewritings could exist, however, their proposed sample-based approximate approach finds minimal rewriting of the original query.
Queries are transformed into a canonical form as a preprocessing step. Next, the search space of potential rewritings is discretized, in such an order that an approximation of the optimal solution can be determined in the next step, by inspecting the succeeding finite set of points. The modified input query meeting coverage constraints can be acquired by examining the grid resulting from the preprocessing step, in an order that ensures the fast identification of the closest rewriting, and by confirming constraint satisfaction using a sample-based approach. The coverage-based rewriting is approximate as a result of the discretization of the search space and of the error in estimating cardinalities and constraint satisfaction on the sample. 
They propose 3 algorithms including a baseline for coverage-based query rewriting. Coverage-based Rewriting Baseline \textit{CRBase} visits the grid in increasing order of distance from the first cell of the grid. During the visit, we look for the cell corresponding to the query with the minimum cardinality that satisfies coverage-based constraints. \textit{CRBase with Pruning (CRBaseP)} adds some pruning rules to reduce the search space and \textit{CRBase with Pruning with Iteration (CRBasePI)} further optimizes \textit{CRBaseP} by iteratively increasing the number of bins during the search up to a given maximum. As a result, each iteration increases the precision by which they refine the query and compute the cardinalities.
Finally, let us demonstrate how CRBase algorithm operates on our running example:
\begin{example}[CRBase (Accinelli et al. 2020)]
Recall data set $\dee$ from section \ref{sec:running-example}. Consider a simple classification task of whether or not an employee makes greater than 50K a year on individuals working more than 40 hours a week and having more than 5 years of experience. Suppose that the selection conditions hours-per-week>40 and years-experience>5 lead to an imbalance in the resulting data set with 130 {\tt single} and 13 {\tt married} individuals while at least 70 of each group is needed. 
Therefore the query needs to be rewritten such that sufficient coverage for the {\tt married} group is met. The algorithm initially transforms the selection conditions into canonical form -hours-per-week<-40 and -years-experience<-5. The search space of interest is now -hours-per-week>-40 and -years-experience>-5 and the goal is to find the closest point to $Q(-5,-40)$ such that the cardinality of {\tt married} individuals is greater than 70. To do so, the algorithm performs an equi-depth (or equi-width) binning with 4 bins on each dimension in the search space. In the resulting grid, each of the grid points represents an SPJ sensitive query obtained from $Q$ by replacing selection constants with the grid point coordinates. Starting from $Q$, the grid points are traversed with various strategies until a point at the minimum distance from $Q$ that meets the coverage condition is found. Suppose that this point is $Q^{\prime}(-4.5, -36)$, therefore the query is re-written as hours-per-week>36 and years-experience>4.5.
\end{example}

Since the proposed methods in \cite{accinelli2020coverage} are approximate, Accinelli et al. further expand their approach in \cite{accinelli2021impact} by introducing some measures for computing the appearing errors. These errors include approximation error resulting from the usage of the grid for the discretization of the query search space, the approximation error correlated with the usage of a sample during the preprocessing and processing phases, and finally, the error related to the detected optimal rewriting.

As a continuum to the previous works in \cite{accinelli2020coverage, accinelli2021impact}, Accinelli et al. \cite{accinelli2022coverage} further extend the considered queries and constraints and also the proposed accuracy measures.

Shetiya et al. \cite{shetiya2022fairness} propose a fairness-aware query rewriting approach in range queries. They use representation ratio as their measure of fairness to address selection bias and try to rewrite the original query such that the most similar results to the original query are returned while meeting the fairness criteria. Depending on the number of predicates in the query, they propose three algorithms. First, \textit{Single Predicate Query Answering (SPQA)} algorithm for single predicate range queries benefits from index jump pointers and quickly looks up fair ranges that have a similarity of more than a threshold. Jump pointers are linear-size indices that enable sub-linear query answering time. Let us demonstrate how SPQA works using our running example:
\begin{example}[SPQA (Shetiya et al. 2021)]
Consider data set $\dee$ from Section \ref{sec:running-example}. Suppose that we are interested in finding individuals who work greater than 40 hours a week. By performing a selection query on the data set, suppose that we observe a 20\% difference between the number of {\tt male} and {\tt female} entities in the query outcome. Considering gender equity, we want to have at most a 5\% difference between the number of {\tt male} and {\tt female} individuals. SPQA finds the most similar fair range to the input query by moving along a jump pointer. Initially, the start end-point of the range is fixed and SPQA
expands the end end-point until a fair query is found. When the window indicating the start and end of the fair range is swept to the left, the start end-point can perform a
shrink or an expansion.
Finally, a fair query (38<hours-per-week<44) most similar to the original query is determined such that the difference in the number of {\tt male} and {\tt female} individuals is less than 5\% while the Jaccard similarity between the two queries results is $\simeq$80\%.
\end{example}
\textit{Best First Search Multi-Predicate (BFSMP)} algorithm models the problem of multi-predicate query answering as the traversal over a graph where nodes represent different queries and there is an edge between two nodes if their outputs differ by one tuple. Starting from the input range, \textit{BFSMP} efficiently explores neighboring nodes to find the most similar fair range. Finally, inspired by the A* algorithm, they propose \textit{Informed BFSMP (IBFSMP)} which improves \textit{BFSMP} using an upper bound on the Jaccard similarity for effective graph exploration.

Moskovitch et al.\cite{moskovitch2022bias} propose using the notion of provenance to mitigate bias in databases by finding minimal query relaxations that increase the number of tuples in groups satisfying a predicate. To do so, the tuples in the data set are annotated with the query selection conditions, and annotations are propagated in the query evaluation phase. The annotated provenance value for each tuple is \textit{prov(t) = }$\prod_{i=1}^{i=k}A_{i[t.A_i]}$ and the provenance inequality of the interested constraint is $Q(D)_{\mathcal{G}}=\sum_{t \in Q(D)_{\mathcal{G}}} prov(t) \geq x$. If the provenance inequality holds for a query then the truth of the quality $\mathcal{T_P}(p)$ is \textit{true}. Next, using the provenance inequality, they present a method for generating minimal relaxations. They use a minimal changes table (MCT) with values being the terms in the provenance inequality sorted in ascending order by their minimal change w.r.t. each column. Finally, they traverse the table in a left-right top-down fashion and keep a result set which they add relaxations or remove them from.

\begin{example}[Query Relaxation (Moskovitch et al. 2022)]
Consider the Example in Section \ref{sec:running-example}, suppose a query wants to select some people who are aged over 60. The fairness requirement is that the results should contain more than 5 {\tt Black female} aged over 60. However, the query result only gives 3 records satisfying the condition. Query relaxation is used to relax the predicates on the continuous values in the query to include more entities in the result. A minimal relaxation is one that no other query relaxation returns a subset of it. For example, changing the search condition from {\tt Black female} aged over 60 to {\tt Black female} aged over 50 could be a minimal relaxation to get enough records if no other query relaxation on the age attribute is closer to the original query and satisfies the fairness requirement.
\end{example}

\subsection{Summary}
Finally, in Figure \ref{fig:table_structured}, we present an overview of the algorithms/techniques described in this section and present a side-by-side comparison between them based on different properties. Each technique is associated with its reference paper and is examined based on the following properties:
\begin{itemize}
    \item \textit{Attribute Type} specifies whether the data is in discrete or continuous space.
    \item \textit{Relation Model} specifies whether data is in single or multiple tables.
    \item \textit{Task} specifies whether the algorithm identifies or resolves insufficient representation.   
    \item \textit{Technique} briefly mentions the general idea of the proposed approach.
\end{itemize}

\newgeometry{left=2cm,bottom=1cm}
\begin{landscape}
\begin{figure}
    \footnotesize
    \centering
    \def\arraystretch{1.5}
    \begin{tabular}{||c|c|c|c|c||}\hline
    
        Algorithm / Method&Attribute Type&Relation Model&Task&Technique\\  
        &\begin{tabular}{C{1.0cm}|C{1.2cm}}Discrete&Continuous\end{tabular}
        &\begin{tabular}{C{0.7cm}|C{0.5cm}}Single&Multi\end{tabular}
        && \\ [0.5ex] \hline \hline
        
        Pattern-Breaker, Pattern-Combiner, Deep-Diver \cite{asudeh2019assessing}&
        \begin{tabular}{C{0.5cm}|C{0.5cm}}\checkmark&\end{tabular}&
        \begin{tabular}{C{0.5cm}|C{0.5cm}}\checkmark&\end{tabular}&
        Identification&
        Pruning descendants and/or ancestors of largest uncovered patterns\\
        \hline
        Decision Tree Training \cite{chung2019slice}&
        \begin{tabular}{C{0.5cm}|C{0.5cm}}\checkmark&\checkmark\end{tabular}&
        \begin{tabular}{C{0.5cm}|C{0.5cm}}\checkmark&\end{tabular}&
        Identification&
        BFS over the ordered decision tree nodes optimized on the classification results\\
        \hline
        
        Lattice Searching \cite{chung2019slice}&
        \begin{tabular}{C{0.5cm}|C{0.5cm}}\checkmark&\end{tabular}&
        \begin{tabular}{C{0.5cm}|C{0.5cm}}\checkmark&\end{tabular}&
        Identification&
        BFS over the lattice of all data slices\\
        \hline

        Generate Top-k Explanations \cite{pradhan2021interpretable}&
        \begin{tabular}{C{0.5cm}|C{0.5cm}}\checkmark&\end{tabular}&
        \begin{tabular}{C{0.5cm}|C{0.5cm}}\checkmark&\end{tabular}&
        Identification&
        Using interventions to measure the effect of biased patterns\\
        \hline

        FunctionAl dependencIes to discoveR Data Bias
        \cite{azzalini2021fair,azzalini2021short}&
        \begin{tabular}{C{0.5cm}|C{0.5cm}}\checkmark&\end{tabular}&
        \begin{tabular}{C{0.5cm}|C{0.5cm}}\checkmark&\end{tabular}&
        Identification&
        Investigating conditional functional dependencies causing discrimination\\
        \hline
        
        SliceLine Enumeration Algorithm \cite{sagadeeva2021sliceline}&
        \begin{tabular}{C{0.5cm}|C{0.5cm}}\checkmark&\end{tabular}&
        \begin{tabular}{C{0.5cm}|C{0.5cm}}\checkmark&\end{tabular}&
        Identification&
        Improved exact sparse linear-algebra implementation of slice enumeration algorithm\\
        \hline
        
        COUNTATA \cite{10.14778/3415478.3415486}, Pattern Count-based Labels  \cite{Moskovitch2021PatternsCL}&
        \begin{tabular}{C{0.5cm}|C{0.5cm}}\checkmark&\end{tabular}&
        \begin{tabular}{C{0.5cm}|C{0.5cm}}\checkmark&\end{tabular}&
        Identification&
        Labels of limited size to estimate the counts of patterns. \\
        \hline
        
        P-Walk \cite{lin2020identifying}&
        \begin{tabular}{C{0.5cm}|C{0.5cm}}\checkmark&\end{tabular}&
        \begin{tabular}{C{0.5cm}|C{0.5cm}}&\checkmark\end{tabular}&
        Identification&
        Priority-based algorithm to improve pruning efficiency of coverage analysis\\
        \hline
        
        DivExplorer \cite{pastor2021looking}&
        \begin{tabular}{C{0.5cm}|C{0.5cm}}\checkmark&\end{tabular}&
        \begin{tabular}{C{0.5cm}|C{0.5cm}}\checkmark&\end{tabular}&
        Identification&
        Exploring large problematic data slices based on divergence\\
        \hline
        
        Shapley Slice Ranking Mechanism \cite{farchi2021ranking}&
        \begin{tabular}{C{0.5cm}|C{0.5cm}}\checkmark&\end{tabular}&
        \begin{tabular}{C{0.5cm}|C{0.5cm}}\checkmark&\end{tabular}&
        Identification&
        Ranking data slices based on Shapley value\\
        \hline
        
        FairVis \cite{cabrera2019fairvis}&
        \begin{tabular}{C{0.5cm}|C{0.5cm}}\checkmark&\end{tabular}&
        \begin{tabular}{C{0.5cm}|C{0.5cm}}\checkmark&\end{tabular}&
        Identification&
        Clustering data set to find problematic subgroups\\
        \hline
        
        Uncovered-2D \cite{asudeh2021coverage}&
        \begin{tabular}{C{0.5cm}|C{0.5cm}}&\checkmark\end{tabular}&
        \begin{tabular}{C{0.5cm}|C{0.5cm}}\checkmark&\end{tabular}&
        Identification&
        Using $k$-th Voronoi diagrams to validate the coverage of a query point\\
        \hline
        
        Uncovered-MD \cite{asudeh2021coverage}&
        \begin{tabular}{C{0.5cm}|C{0.5cm}}&\checkmark\end{tabular}&
        \begin{tabular}{C{0.5cm}|C{0.5cm}}\checkmark&\end{tabular}&
        Identification&
        Creating random samples and learning the uncovered using \enet approximation\\
        \hline
        
        Iterative Algorithm for Slice Tuner \cite{tae2021slice}&
        \begin{tabular}{C{0.5cm}|C{0.5cm}}&\end{tabular}&
        \begin{tabular}{C{0.5cm}|C{0.5cm}}\checkmark&\end{tabular}&
        Resolution&
        \pbox{7cm}{Periodically updating learning curves for dependent slices to learn the amount of data needed to be collected}\\
        \hline

        Greedy Coverage Enhancement \cite{asudeh2019assessing}&
        \begin{tabular}{C{0.5cm}|C{0.5cm}}\checkmark&\end{tabular}&
        \begin{tabular}{C{0.5cm}|C{0.5cm}}\checkmark&\end{tabular}&
        Resolution&
        Transformation to Hitting-Set problem to collect minimum required data\\
        \hline
        
        Modified Greedy Hit Count \cite{azzalini2021functional}&
        \begin{tabular}{C{0.5cm}|C{0.5cm}}\checkmark&\end{tabular}&
        \begin{tabular}{C{0.5cm}|C{0.5cm}}\checkmark&\end{tabular}&
        Resolution&
        Transformation to Hitting-Set problem to collect minimum required data\\
        \hline
        
        SMOTE \cite{DBLP:journals/jair/ChawlaBHK02}&
        \begin{tabular}{C{0.5cm}|C{0.5cm}}\checkmark&\end{tabular}&
        \begin{tabular}{C{0.5cm}|C{0.5cm}}\checkmark&\end{tabular}&
        Resolution&
        Over-sampling minority group instances\\
        \hline
        
        Greedy Fairness-aware Data Augmentation \cite{sharma2020data}&
        \begin{tabular}{C{0.5cm}|C{0.5cm}}\checkmark&\end{tabular}&
        \begin{tabular}{C{0.5cm}|C{0.5cm}}\checkmark&\end{tabular}&
        Resolution&
        Augmenting minority group instances to reach the ideal data set\\
        \hline
        
        Re-weighting \cite{celis2020data}&
        \begin{tabular}{C{0.5cm}|C{0.5cm}}&\checkmark\end{tabular}&
        \begin{tabular}{C{0.5cm}|C{0.5cm}}\checkmark&\end{tabular}&
        Resolution&
        Re-weighting data to meet representation constraints \\
        \hline

        \pbox{6cm}{Coverage-based Rewriting Baseline Algorithm with Pruning (with Iteration) \cite{accinelli2020coverage,accinelli2022coverage,accinelli2021impact}}&
        \begin{tabular}{C{0.5cm}|C{0.5cm}}\checkmark&\checkmark\end{tabular}&
        \begin{tabular}{C{0.5cm}|C{0.5cm}}\checkmark&\end{tabular}&
        Resolution&
        \pbox{7cm}{Query rewriting such that representation constraints are met}\\
        \hline
        
        \pbox{6cm}{Single-predicate range queries answering, (Informed) Best first search multiple-predicate\cite{shetiya2022fairness}}&
        \begin{tabular}{C{0.5cm}|C{0.5cm}}&\checkmark\end{tabular}&
        \begin{tabular}{C{0.5cm}|C{0.5cm}}\checkmark&\end{tabular}&
        Resolution&
        \pbox{7cm}{Query rewriting such that representation constraints are met}\\
        \hline
        
        Modified threshold algorithm \cite{moskovitch2022bias}&
        \begin{tabular}{C{0.5cm}|C{0.5cm}}&\checkmark\end{tabular}&
        \begin{tabular}{C{0.5cm}|C{0.5cm}}\checkmark&\end{tabular}&
        Resolution&
        Minimal query relaxations based on provenance \\
        \hline
        
        \pbox{5cm}{Dynamic Programming Algorithm \\ for Data Distribution Tailoring}\cite{nargesian2021tailoring}&
        \begin{tabular}{C{0.5cm}|C{0.5cm}}\checkmark&\end{tabular}&
        \begin{tabular}{C{0.5cm}|C{0.5cm}}&\checkmark\end{tabular}&
        Resolution&
        Exact Dynamic Programming to integrate data from multiple sources \\
        \hline
        
        Equi-cost Binary \cite{nargesian2021tailoring}&
        \begin{tabular}{C{0.5cm}|C{0.5cm}}\checkmark&\end{tabular}&
        \begin{tabular}{C{0.5cm}|C{0.5cm}}&\checkmark\end{tabular}&
        Resolution&
        \pbox{7cm}{Data integration for binary valued attribute with equal data collection cost}\\
        \hline
        
        Coupon Collector \cite{nargesian2021tailoring}&
        \begin{tabular}{C{0.5cm}|C{0.5cm}}\checkmark&\end{tabular}&
        \begin{tabular}{C{0.5cm}|C{0.5cm}}&\checkmark\end{tabular}&
        Resolution&
        Data integration cost approximation via instances of coupon collector \\
        \hline
        
        Upper Confidence Bound \cite{nargesian2021tailoring}&
        \begin{tabular}{C{0.5cm}|C{0.5cm}}\checkmark&\end{tabular}&
        \begin{tabular}{C{0.5cm}|C{0.5cm}}&\checkmark\end{tabular}&
        Resolution&
        \pbox{7cm}{Data integration modeled as a multi-armed bandit when data source distributions are unknown} \\
        \hline

        Min-max Stochastic Gradient Descent
        \cite{abernethy2020adaptive,abernethy2020active}&
        \begin{tabular}{C{0.5cm}|C{0.5cm}}\checkmark&\end{tabular}&
        \begin{tabular}{C{0.5cm}|C{0.5cm}}&\checkmark\end{tabular}&
        Resolution&
        Adaptive sampling approach to data integration\\
        \hline

        Optimistic Sampling for Fair Classification
        \cite{shekhar2021adaptive}&
        \begin{tabular}{C{0.5cm}|C{0.5cm}}\checkmark&\end{tabular}&
        \begin{tabular}{C{0.5cm}|C{0.5cm}}&\checkmark\end{tabular}&
        Resolution&
        \pbox{7cm}{Adaptive sampling modeled as a multi-armed bandit for data integration}\\
        \hline

        \pbox{6cm}{Uniform, Lower Upper Confidence Bound, Thompson Sampling
        \cite{niss2022achieving}}&
        \begin{tabular}{C{0.5cm}|C{0.5cm}}\checkmark&\end{tabular}&
        \begin{tabular}{C{0.5cm}|C{0.5cm}}&\checkmark\end{tabular}&
        Resolution&
        \pbox{7cm}{Reducing adaptive sampling to convex feasibility problem to check the feasibility of data integration}\\
        \hline
        
        Query-2D \cite{asudeh2021coverage}&
        \begin{tabular}{C{0.5cm}|C{0.5cm}}&\checkmark\end{tabular}&
        \begin{tabular}{C{0.5cm}|C{0.5cm}}\checkmark&\end{tabular}&
        Resolution&
        Generating a signal of whether query results can be trusted or not \\
        \hline
        
        Query-MD \cite{asudeh2021coverage}&
        \begin{tabular}{C{0.5cm}|C{0.5cm}}&\checkmark\end{tabular}&
        \begin{tabular}{C{0.5cm}|C{0.5cm}}\checkmark&\end{tabular}&
        Resolution&
        Generating a signal of whether query results can be trusted or not \\
        \hline
        
        Datasheets for Data sets \cite{gebru2018datasheets}, MithraLabel \cite{sun2019mithralabel}&
        \begin{tabular}{C{0.5cm}|C{0.5cm}}\checkmark&\checkmark\end{tabular}&
        \begin{tabular}{C{0.5cm}|C{0.5cm}}\checkmark&\checkmark\end{tabular}&
        Resolution&
        Describes data sets from representation perspectives \\
        \hline
        
    \end{tabular}
    \caption{Properties of different techniques for identifying and resolving representation bias in structured data.}
    \label{fig:table_structured}
\end{figure}
\end{landscape}
\restoregeometry

\section{Representation Bias in Unstructured Data} 
\label{sec:unstructured}
There has been extensive work on techniques for identifying and resolving representation bias in tabular data sets, as we have discussed above. Additionally, there also is research investigating representation concerns in unstructured data types such as images, text, and graphs.
In this section, we discuss the body of literature on identifying and mitigating representation bias in unstructured data.

\subsection{Representation Bias in Image Data}

\begin{figure}[!h]
    \footnotesize
    \centering
        \begin{tikzpicture}[
            level 1/.style={sibling distance=7cm},
            level 2/.style={sibling distance=2cm},
            level 3/.style={sibling distance=2cm}]
            \node {\textbf{Representation Bias in Image Data}}
                child {node {Identification}
                    child {node[align=center] {Reduction to \\Tabular Data \\\cite{dulhanty2019auditing,buolamwini2018gender,merler2019diversity,wang2020revise}}
                    }
                    child {node[align=center] {Biased Image \\Representations \\\cite{karkkainen2021fairface}}
                    }
                    child {node[align=center] {Cross Data Set \\Bias Detection\\\cite{torralba2011unbiased,khosla2012undoing,schaaf2021towards}}
                    }
                    child {node[align=center] {Crowd-sourcing\\\cite{hu2020crowdsourcing}}
                    }
                    }
                child {node {Resolution}
                    child {node[align=center] {Data Augmentation\\\cite{jaipuria2020deflating,georgopoulos2021mitigating,yucer2020exploring,goel2020model}}}
                    child {node[align=center] {Reweighting\\\cite{li2019repair}}}
                };
        \end{tikzpicture}
        \caption{Classification of techniques on identifying and resolving representation bias in image data sets}
\label{fig:taxonomy_image}
    \end{figure}
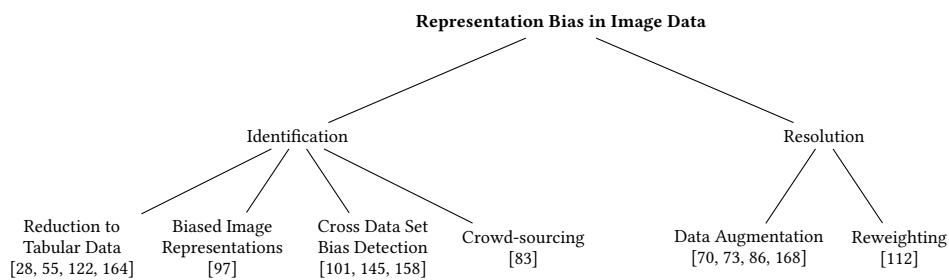

Computer vision systems have recently achieved outstanding capacity. 
Identification and resolution of unwanted biases, specifically the ones due to the disproportionate representation in the image data sets, have drawn a lot of attention from different research communities. In this section, inspired by Fabbrizzi et al. \cite{fabbrizzi2021survey}, we present a taxonomy (as seen in Figure \ref{fig:taxonomy_image}) to classify the techniques and followed by its structure, we review the techniques for debiasing image data sets. Additionally, while the extent of the works studied in this section is broader than those reviewed in \cite{fabbrizzi2021survey}, we would like to direct the interested reader to \cite{fabbrizzi2021survey} for a more comprehensive survey exclusively on the subject.

\subsubsection{Identification of Representation Bias}
    
\paragraph{\textbf{Reduction to tabular data}}
    
The main idea of this group of techniques is to transform the image data into tabular data to benefit from the rich literature on the identification of bias in tabular data. The transformation process involves direct feature extraction from the images using recognition tools and/or indirectly using the metadata of the image such as description, tag, etc. Of course, these automatic techniques may themselves perpetuate and amplify the biases in the data as they are prone to errors.
In this regard, Dulhanty et al. \cite{dulhanty2019auditing} evaluate two subsets of ImageNet \cite{deng2009imagenet} with human images for representation bias w.r.t. gender and age. They first apply a face recognition algorithm to the data and next, they apply gender and age recognition models to the outcome. With age and gender attributes determined, they calculate the distributions among genders and age groups.
    
Buolamwini et al.\cite{buolamwini2018gender} created a benchmark data set with balanced entities w.r.t. gender and skin color by counting and used it to audit the existing gender classification models.
    
Merler et al. \cite{merler2019diversity}, propose utilizing information theoretical measures of diversity and evenness such as Shannon entropy, Simpson index, etc. to construct balanced data sets. However, they use existing recognition models or annotators for labeling the images w.r.t. gender and race.
    
Wang et al. \cite{wang2020revise} build a tool named REVISE for identifying and mitigating bias in visual data sets. Their scope is limited to three sets of metrics: 1) \textit{Object-based} that focuses on statistics about object frequency, scale, context, or diversity of representation 2) \textit{Person-based} that examines the representation of people from various demographics in the data set, and allows the user to assess what potential downstream consequences this may have to consider how best to intervene. It also builds on the object-based analysis by considering how the representation of objects with people of different demographic groups differs. 3) \textit{Geography-based} that considers the portrayal of different geographic regions within the data set and is deeply intertwined with the previous two, as geography influences both the types of objects that are represented, as well as the different people that are pictured. REVISE accepts annotated image data sets as input and depending on the annotations it provides insights on the data sets based on each of the three categories of metrics explained above. Metrics such as object count, scale, co-occurrence, scene diversity, etc. for \textit{Object-based} category, person prominence, appearance differences, and contextual representations for \textit{Person-based} and geography distributions based on people, language, weather and etc. for \textit{Geography-based} category. REVISE does not claim to find all the visual biases and it is limited to the available annotations accompanying the data.
    
\paragraph{\textbf{Biased Image Representations}}
The techniques in this group use distance-based analysis on the low-dimensional representation of the images in the embedding space to identify representation bias. 
Particularly, Karkkainen et al. \cite{karkkainen2021fairface} create a balanced face data set w.r.t. age, race, and gender. To evaluate the diversity of their data set compared to the existing work, they visualize the images in 2D using t-SNE \cite{van2008visualizing}, a statistical method for visualizing high dimensional data by giving each data point a location in 2D/3D space, on the embeddings trained on multiple online sources. Next, they measure pairwise Manhattan distances between random subsets of the images based on their 128-dimensional embedding. The skewness of the resulting distribution towards high distances is evidence of high diversity and proper representation of different subgroups.
    
\paragraph{\textbf{Cross Data Set Bias Detection}}
Each data set includes specific \textit{signature} biases that make it distinct from the rest. This signature bias is introduced in the data collection process and affects the generalizability of the models built on the data set. This group of methods evaluates the signature bias by comparing different data sets.
     
In this regard, Torralba et al.\cite{torralba2011unbiased} perform some experiments on famous image data sets to measure the bias. To correctly measure the bias of a data set, it should be compared to the real visual world, which would have to be in the form of a data set, which could also be biased and, consequently, not a viable option. Therefore, they suggest \textit{Cross-data set Generalization} by training a model on a data set and testing it on another. Assuming that the training data set is truly representative of the real world, the model should perform well; otherwise, it means that there are biases, such as selection and capture, present in the data set. Next, knowing that data sets define a visual phenomenon not only by what it is but also by what it is not, they argue about \textit{Negative Set Bias} and whether the negative samples are representative of the rest of the world or even sufficient. To do so, they run an experiment such that for each data set, a classifier is trained on its own set of positive and negative instances, and then during testing, the positives come from that data set, but the negatives come from all data sets combined. The performance of the models shows how well the data set is representing the rest of the world. 
    
Khosla et al. \cite{khosla2012undoing} propose an algorithm that learns the \textit{visual world} model and the biases for each data set. The key observation is that all data sets are sampled from a common \textit{visual world} (a more general data set). A model trained on this data set would have the best generalization ability, however, making such a data set is not realistic. Therefore, they suggest defining the biases associated with each data set and approximating the weights for the visual world by removing the bias from each data set. The visual world model performs well on average but is not necessarily the best on any specific data set since it is not biased towards any one data set. On the other hand, the biased model, built by combining the visual world model and the learned bias, performs superior on the data set that it is biased towards but does not necessarily extend to the rest of the data sets. In this regard, they propose a maxed-margin learning discriminative framework to collectively learn the weight vector correlated to the visual world object model and a set of bias vectors, for each data set such that when combined with the visual world weights lead to an object model specific to the data set.
    
Another related work by Schaaf et al.~\cite{schaaf2021towards} focuses on measuring bias in image classification tasks by means of attribution maps. Attribution maps seek to explain image classification models, such as CNNs, by demonstrating the importance of each individual pixel of the input image on the outcome. To do so, they propose a four-step process to indicate their usefulness. First, they generate artificial data sets with a known bias. For example, they generate a biased fruit data set where apples are all on tree backgrounds, while other fruits have different backgrounds, and an unbiased data set where all fruits have different backgrounds. Next, they train biased CNN models and then generate attribute maps using different attribution techniques such as \textit{Grad-CAM}, \textit{Score-CAM}, \textit{Integrated Gradients} and \textit{epsilon-LRP}. Finally, they quantitatively evaluate attribution maps' ability to detect bias using metrics such as \textit{Relevance Mass Accuracy (RMA)}, \textit{Relevance Rank Accuracy (RRA)} and \textit{Area Over The Perturbation Curve (AOPC)}. Their results partly confirm the ability of attribution maps to quantify bias. However, in some cases, attribution maps provide inconsistent results for different metrics.
    
\paragraph{\textbf{Crowd-sourcing}}
Hu et al.~\cite{hu2020crowdsourcing} propose a crowd-sourcing workflow to facilitate sampling bias discovery in visual data sets with the help of human-in-the-loop. This workflow takes a visual data set as input and outputs a list of potential biases of the data set. There are three steps in this workflow. The first step is \textit{Question Generation} in which the crowd inspects random samples of images from the input data set and describes their similarity using a question-answer pair. The next step is \textit{Answer Collection} in which the crowd reviews separate random samples of images from the input data set and provides answers to questions generated in the earlier step. Finally, in the third step called \textit{Bias Judgement}, the crowd judges if the statements about the visual data set automatically generated through the accurate questions and answers collected in the former steps reflect the real world.
    
\subsubsection{Resolving Representation Bias}
\paragraph{\textbf{Data Augmentation}}
This group of techniques tries to mitigate bias by adding samples for the underrepresented groups benefiting from the rich literature on image augmentation.
    
Jaipuria et al. \cite{jaipuria2020deflating} propose a bias mitigation approach by using targeted synthetic data augmentation that combines the advantages of gaming engine simulations and \textit{sim2real} style transfer techniques to bridge the gaps in real data sets for vision tasks. However, instead of blindly collecting more data or mixing data sets that often end up in worse final performance, they suggest a smarter approach to augment data regarding the task-specific noise factors. The results consistently indicate that through adding synthetic data to the training set, a noticeable improvement occurs in cross-data set generalization, in contrast, to merely training on original data, for a training set of equal size.
    
Georgopoulos et al. \cite{georgopoulos2021mitigating} propose a style transfer approach based on generative adversarial networks (GANs), capable of creating additional images, reflecting multiple attributes such as race, gender, and age. The resulting data set is less biased w.r.t. the aforementioned attributes. This is accomplished by relaxing the strict reliance on a single attribute label and adding a tensor-based mixing structure that multilinearly represents multiplicative interactions between attributes.
    
Similarly, Yucer et al. \cite{yucer2020exploring} propose another adversarial augmentation method utilizing CycleGANs to transfer race to mitigate representation bias. They aim to create a synthesized data set by transforming facial images into different racial domains while maintaining identity-related traits so that race-related traits eventually become irrelevant in determining the subject's identity.
    
Goel et al. \cite{goel2020model} propose an advanced augmentation approach that is oblivious to the differences within subgroups and aims for class information shared by subgroups. In this regard, they propose CycleGAN Augmented Model Patching (CAMEL) that first, learns mappings between pairs of subgroups using CycleGANs and creates transformations that can be used to generate augmented examples based on the training instances and second leverages the transformations as data augmentations and builds a more robust classifier.
    
\paragraph{\textbf{Reweighting}}
Li et al.~\cite{li2019repair} propose \textit{REPAIR}, a resampling-based bias mitigation approach that is formulated as an optimization problem. \textit{REPAIR} assigns a weight to the instances that the classifier built on a feature representation can penalize more easily. This is implemented through a deep neural network as a feature extractor for the representation of interest and learning an independent linear classifier to classify the extracted features. Next, bias mitigation is defined as maximizing the ratio between the loss of the classifier on the reweighted data set and the uncertainty of the ground-truth labels. Lastly, the problem is reduced to a minimax problem, which can be solved by alternatingly updating the classifier coefficients and the data set resampling weights, through stochastic gradient descent.

\subsection{Representation Bias in Natural Language Data}
\begin{figure}[!h]
    \footnotesize
    \centering
        \begin{tikzpicture}[
            level 1/.style={sibling distance=7.5cm},
            level 2/.style={sibling distance=4.5cm},
            level 3/.style={sibling distance=2.3cm}]
            \node {\textbf{Representation Bias in Textual Data}}
                child {node {Identification}
                    child {node[align=center] {Performance and Representation \\Difference among Sensitive Groups \\\cite{dixon2018measuring,badjatiya2019stereotypical}}
                    }
                    child {node[align=center] {Analyzing Sub-space \\Embeddings of Sensitive Attribute\\\cite{bolukbasi2016man,manzini2019black,papakyriakopoulos2020bias}}
                    }
                    }
                child {node {Resolution}
                    child {node[align=center] {Debiasing Training \\Corpora}
                        child {node[align=center] {Data Augmentation\\\cite{lucy2021gender,zhao2018gender,vanmassenhove2019getting}}}
                        child {node[align=center] {Bias fine-tuning\\\cite{park2018reducing}}}
                    }
                    child {node[align=center] {Debiasing Embeddings}
                        child {node[align=center] {Removing Sub-space \\of Sensitive Attribute\\\cite{schmidt2015rejecting,bolukbasi2016man,manzini2019black,chen2018my}}}
                        child {node[align=center] {Learning Neutral \\Embeddings\\\cite{zhao2018learning}}}
                    }
                };
        \end{tikzpicture}
        \caption{Classification of techniques on identifying and resolving representation bias in textual data}
\label{fig:taxonomy_nlp}
    \end{figure}
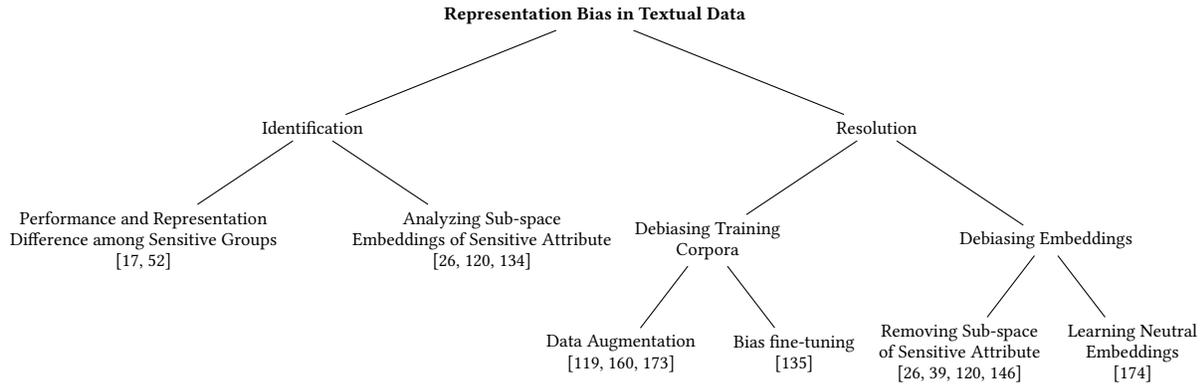

Natural language processing (NLP) is one of the areas that has widely been affected by the data explosion and advancement of data-driven decision-making systems. However, the existing biases in the data have regularly resulted in discriminatory outcomes w.r.t. gender, race, age, disability, etc. Representation bias as one of the key reasons for such issues has been extensively studied in different NLP tasks such as machine translation, caption generation, sentiment analysis, hate speech detection, coreference resolution, language models, and word embeddings. Hundreds of technical papers with a variety of solutions and dozens of reviews have been published tackling different angles of the matter w.r.t. the task and the target of the bias. Going through, the details of each work is out of the scope of this survey due to the richness of existing surveys \cite{sun2019mitigating, blodgett2020language, garg2022handling, savoldi2021gender,diaz2018addressing,venkit2021identification}, however, we try to give an overview and a taxonomy of the techniques (as seen in Figure \ref{fig:taxonomy_nlp}) on identifying and mitigating representation bias in textual data while giving proper directions to the curious reader.

Representation bias in textual data can happen as a result of the following \cite{sun2019mitigating}:
\begin{itemize}
    \item \textit{Denigration:} Using culturally or historically derogatory words.
    \item \textit{Stereotyping:} Heightening the existing societal stereotypes. 
    \item \textit{Under-representation:} Disproportionately low representation of a specific group.
\end{itemize}
Each NLP task can be associated with one or more of these classes as demonstrated in \cite{sun2019mitigating}. Next, inspired by \cite{sun2019mitigating}, we present a taxonomy for the classification of techniques for identifying and mitigating representation bias in textual data, and following the structure of the taxonomy, we provide a summary of the techniques in the latter sections.

\subsubsection{Identification of Representation Bias}
There are two major approaches for identifying representation bias in the NLP literature:
\paragraph{\textbf{Performance and Representation Difference Among Sensitive Groups}} Regardless of the task, most NLP model predictions should not be significantly affected by a sensitive attribute such as gender, race, etc. of the entity. Following this fact and regarding representation bias in the context of gender, gender swapping and measuring the difference in evaluation score (such as false-positive rate difference or false-negative rate difference) is a common practice to assess gender bias in such tasks. Furthermore, standard evaluation data sets commonly used in NLP are not sufficient for measuring gender bias as they often contain bias themselves due to the disproportionate representation of male and female entities. Therefore, carefully designed task-specific data sets known as Gender Bias Evaluation Test Sets (GBETs) are constructed that can control the effect of gender bias. 

Aside from the performance aspect, Dixon et al. \cite{dixon2018measuring} show how imbalances in the training data w.r.t. representation can lead to biases in the constructed text classification models with potentially unfair results towards the under-represented group. 
An example of such biases can be seen in toxicity detection models where due to disproportionate representation of terms such as ``gay'' in the training data, statements such as ``I'm a gay man'' are assigned overly high toxicity scores even though the comment is not toxic. 
Models are falsely biased toward words that are disproportionately represented in toxic comments compared to the overall data set and also they tend to be more biased toward short comments. To identify the representation bias, Dixon et al. \cite{dixon2018measuring} create a hand-curated list of words for which they study these two properties. Badjatiya et al. \cite{badjatiya2019stereotypical} add two more strategies to what Dixon et al. \cite{dixon2018measuring} proposed to identify representation bias in textual data. The first strategy is investigating skewed occurrences across classes. If a term happens to appear in lots of training samples belonging to the toxic class, it encourages the models to classify a comment containing that particular term as toxic. The second strategy is skewed predicted class probability distribution, which is the maximum probability of a term belonging to a non-neutral class. A high probability value means that the model has stereotyped the term to belong to the toxic/non-toxic class. 

\paragraph{\textbf{Analyzing Sub-space Embeddings of Sensitive Attribute}}
Word embeddings and language models are trained on the available biased text corpora and tend to amplify and propagate these biases to the downstream tasks when used as features.
Bolukbasi et al. \cite{bolukbasi2016man} investigate representation bias in the context of gender in the embedding space by showing that geometrically, gender bias can be captured by a direction. Besides, they show that gender-neutral words (e.g. nurse) are linearly separable from gender-defined words (e.g. queen). Therefore it is possible to differentiate between the two and capture gender bias in the embedding space.
The proposed technique operates as followed:
initially, a set of gender-specific words such as \{he, she, man, woman, \dots\} are chosen as seed words. Using the seed words an SVM classifier is trained to get the rest of the gender-specific words. The complement of the gender-specific corpus grants us the set of gender-neutral words. Having the gender-specific and gender-neutral words separated, they select the seed word pairs such as he-she to act as the x-axis to identify the gender subspace. By checking the distance of gender-neutral words from the he or she end of the axis (``nurse'' closer to she, ``genius'' closer to he), they identify how biased the word embeddings are toward such words. These biases originate from the insufficient association of such words with the opposite gender in the original corpora on which the embeddings were trained. 

Manzini et al. \cite{manzini2019black} extend this solution to non-binary gender and multi-class sensitive attributes such as race. religion, etc.
Papakyriakopoulos et al.\cite{papakyriakopoulos2020bias} study detecting representation bias resulting from historical biases reflected in the word embeddings. To detect the bias of word embeddings, they define an inter-group direction (for instance between man and woman) and then the bias is quantified as the cosine distance between the word vector and the inter-group direction. This method compares the magnitude of dependence between a concept and the two groups. If the concept vector has a higher similarity to a group than another, the concept is considered to be biased in that direction.

\subsubsection{Resolving Representation Bias}
Several methods have been proposed to mitigate representation bias in textual data. Some of these methods require the models to be retrained after the alterations while some do not and only manipulate the model to fix the outcomes. In the following we will introduce each of these methods and reiterate some of the adopted techniques:
\paragraph{\textbf{Text Corpus Alteration}} 
To debias the text corpora, two approaches have been proposed: 
        
\textit{Data Augmentation:} The augmentation approach is to add modified copies of the existing data, or newly created synthetic data, to the corpora.
While some works propose completely removing, masking, or replacing any indication of gender, race, etc. from the text corpora to eliminate representation bias, De Arteaga et al. \cite{de2019bias} makes an interesting observation that even by removing the explicit indicators regarding gender, race, or socioeconomic status in the text corpora, although a slight reduction in representation bias would occur towards the minority group, a significant gap remains due to the imbalances in the available data between the minority and majority group. Similarly, Li et al. \cite{lucy2021gender} make a closely related conclusion for the task of text generation where they investigate representation bias in the stories generated by GPT-3. They demonstrate how gender stereotypes occur in generated narratives, even in the absence of gender indicators or stereotype-related cues. They propose prompt design as a possible workaround for mitigating bias and steering GPT-3, however, they state that it is not a feasible solution for every situation.
Zhao et al. \cite{zhao2018gender} propose another approach to decrease the bias in text corpora by creating an identical but gender-swapped version of the original data set and training the model on the union of the original data set, the gender-swapped version and the named-entity anonymized version of the original data set. 

In tasks such as machine translation, due to the domination of male entities in the available text corpora, the models tend to predict the entities more as male while the actual gender may not be clear. This specifically becomes problematic while translating into languages such as French where words are gender-specific and masking or removal of gender indicators is not an option. Vanmassenhove et al. \cite{vanmassenhove2019getting} propose an augmentation technique known as gender-tagging that tries to solve the aforementioned issue by appending the gender of the entity to the sentences. Gender-tagging preserves the gender of the speaker and therefore, the machine translation model can consider it while making predictions.

\textit{Bias Fine-tuning:} An alternative approach to debias text corpora, proposed by Park et al. \cite{park2018reducing}, is to use transfer learning from an already bias-free data set and fine-tune on the biased data to train a model. This approach enables the models to benefit from bias-free data sets while still sufficiently good to perform the assigned learning task.

\paragraph{\textbf{Word Embedding Adjustment}}
Complete elimination of representation bias from embedding space is not a feasible goal.  However, it has been shown that it is possible to mitigate it w.r.t. the similarity to sensitive attribute subspace and not needing the embeddings to be retrained. To debias the word embeddings, two approaches have been proposed:
        
\textit{Removing Sub-space of Sensitive Attribute:} This is achieved by building a neutral (i.e. genderless, raceless, etc.) framework for all words \cite{schmidt2015rejecting} or for gender-neutral words \cite{bolukbasi2016man}. For instance, Bolukbasi et al. \cite{bolukbasi2016man} propose a neutralization method to debias the word embeddings. Recall that to identify the bias, they projected each gender-neutral word vector on an axis with gender-specific words on each end. Having known that the bias exists, they project the gender-neutral words on the y-axis and thus eliminate the gender bias. Another approach is to make gender-neutral words equidistant to all words in the gender-specific set meaning that the word ``nurse'' will be equidistant to sets \{he, she\} and \{man, woman\}.
Manzini et al. \cite{manzini2019black} show that this solution is extendable to non-binary sensitive attributes.

However, Cheng et al. \cite{cheng2022toward} show that bias w.r.t. different sensitive attributes can be correlated and independent removal of bias may not be sufficient.
To mitigate the bias at a word embedding level, for each bias-sensitive word, they define a sentiment direction by forming pairs showing different ends of bias (e.g. good-bad, positive-negative, etc.) and taking the difference between the word embeddings of words in each set and the mean word embedding over the set. Next, they apply PCA, with the resulting component being the sentiment direction. Next, they define a corresponding set to the neutral words vector (e.g. doctor, nurse, etc.) and hard-neutralize this vector by making it orthogonal to the sentiment vector \cite{bolukbasi2016man}. 

\textit{Learning Neutral Embeddings:} Zhao et al. \cite{zhao2018learning} suggest separating information about the sensitive attribute in a dimension and keeping the neutral information in other dimensions. In doing so, the sensitive attribute information can be utilized or neglected on demand. This method requires retraining the embeddings.

\subsubsection{Representation Bias in Speech Recognition}
Identification and mitigation of representation bias in speech recognition systems have been briefly studied in the contexts of gender, race, and age \cite{koenecke2020racial,feng2021quantifying,liu2022towards}. The primary approach to identifying the bias in such systems is by measuring the error rate of the speech recognition model among different subgroups. Demographic information of the speaker is usually acquired through annotations or utilizing automatic methods \cite{rakesh2011gender,childers1991gender,erokyar2014age}. With the demographic information available, the problem is reduced to bias identification in tabular data. For the purpose of bias mitigation, diversifying the training data sets w.r.t. race, gender, age, etc. through the addition of more data is recommended.

\subsection{Representation Bias in Graphs} 
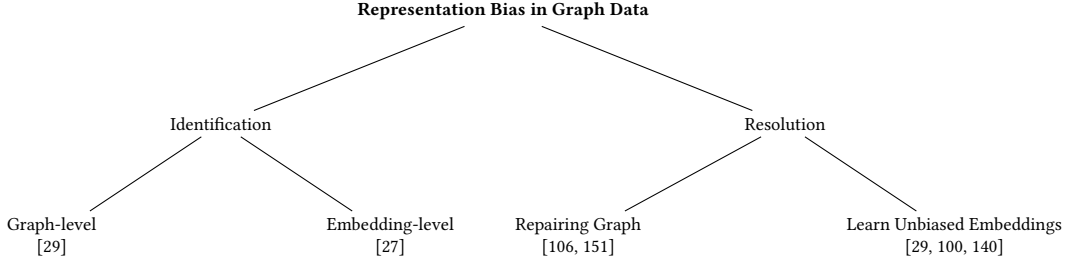
\begin{figure}[!h]
    \footnotesize
    \centering
        \begin{tikzpicture}[
            level 1/.style={sibling distance=7.5cm},
            level 2/.style={sibling distance=4.5cm},
            level 3/.style={sibling distance=2.3cm}]
            \node {\textbf{Representation Bias in Graph Data}}
                child {node {Identification}
                    child {node[align=center] {Graph-level\\\cite{buyl2020debayes}}
                    }
                    child {node[align=center] {Embedding-level\\\cite{bose2019compositional}}
                    }
                    }
                child {node {Resolution}
                    child {node[align=center , xshift = -0.5cm] {Repairing Graph\\\cite{laclau2021all,spinelli2021fairdrop}}}
                    child {node[align=center] {Learn Unbiased Embeddings\\\cite{rahman2019fairwalk,khajehnejad2022crosswalk,buyl2020debayes}}}
                };
        \end{tikzpicture}
        \caption{Classification of techniques on identifying and resolving representation bias in graphs}
\label{fig:taxonomy_graph}
    \end{figure}

The capacity of graphs to model complex phenomena is gaining increasing attention in many domains, including those with high societal impact. The sensitivity of applications such as online polarization, job recommendation systems, disaster response, and criminal justice has led to increasing interest in addressing bias in these systems. There now are comprehensive studies in the form of review papers and tutorials \cite{dong2022fairness,choudhary2022survey,kang2021fair} to identify biases and promote fairness. In this section, inspired by Choudhary et al. \cite{choudhary2022survey}, we discuss recent techniques to identify and mitigate representation bias in graphs, present a taxonomy (as seen in Figure \ref{fig:taxonomy_graph}) of such techniques, and give pointers to the interested reader.

Graphs hold properties such as being non-iid and non-euclidean that make the existing bias identification and mitigation solutions ineffective. The non-iid assumption suggests that an alteration in one node or edge will affect its neighbors in the graph. The non-euclidean assumption states that before performing any learning task, a vectorized representation of the level of interest (node-level, edge-level, or graph-level) should be learned. Aside from the pre-existing bias in the graphs, different objective functions to learn the representations can perpetuate and amplify the biases in the graph embeddings. The embeddings should hold two properties:
\begin{itemize}
    \item They should reflect the properties of the graph structure.
    \item They should be independent of the sensitive attributes.
\end{itemize}
The first property is guaranteed through the choice of the objective function, however, the second property, is our problem of interest and can be secured in a two-staged process of identification and mitigation of bias in a variety of methods.

\subsubsection{Identification of Representation Bias}
This set of methods targets representation bias from two different levels:
    \paragraph{\textbf{Graph-level}} Assortative mixing coefficient \cite{newman2003mixing} is a notion that is used in \cite{buyl2020debayes} to evaluate the homophily of a graph regarding a particular attribute. This notion is used to evaluate the graph structures for the existing biases. The values of the assortative mixing coefficient fall into a range of $[-1,1]$ and the closer the value to $-1$ or $1$, the more correlated the graph is with a sensitive attribute. The mixing coefficient is calculated using the following formula:
    \[
        r=\frac{{\sum_{i}{e_{ij}}}-\sum_{i}{a_i b_i}}{1-\sum_{i}{a_i b_i}}
    \]
    where:
    \[
        e_{ij}=\frac{card\{(i,j) \in \mathcal{E};A_{v_i}=i, A_{v_j}=j\}}{m} \;,\;
        a_i=\sum_j{e_{ij}}\;\text{and}\; b_j=\sum_i{e_{ij}}
    \]
    where $v_i$ is $i$-th vertex, $\mathcal{E}$ is the set of all edges, $m$ is the number of all edges, $A$ is the sensitive attribute, and $a_i$, $b_j$ is the ratio of the edges starting from and ending at each of the attribute values. An $r$ value of zero indicates no bias in the graph. Mixing coefficient value $r$ can be calculated on any graph to determine bias and promote fairness. 
    
    \paragraph{\textbf{Embedding-level}} Representation Bias (RB) \cite{bose2019compositional} (should not be mistaken with the topic of our survey though) refers to the bias in node-level embeddings. RB is calculated using the following:
    \[
        RB=\sum_{a=0}^{l}\frac{1}{{|V_a|}}\text{AUC}(\{\mathbb{P}_h(a,z_v)|\forall v \in V_a\})
    \]
    where $V_a=\{v|A(v)=a\}$ is the set of nodes having sensitive attribute value $a$, $h$ is a classifier trained to predict sensitive attribute $A$ and $\mathbb{P}_h(a,z_v)$ is the result of the classification.
    The idea is to consider the sensitive attribute $A$ as the target variable and then the aforementioned formula calculates the weighted average of the one-vs-rest AUC values from the output of the classifier trained to predict $A$. RB values fall into $[0,1]$ range. The closer the value to 0.5, the more nondiscriminatory the graph is w.r.t. the sensitive attribute. 

\subsubsection{Resolving Representation Bias}
    \paragraph{\textbf{Repairing Graph}}
    The methods introduced in this section try to remove the bias from the graph structure itself rather than the embeddings.
    Laclau et al. \cite{laclau2021all} try to mitigate the bias in the graph structure using optimal transport technique in the context of fair edge prediction. They reduce the problem to the problem of alignment between node distributions of nodes belonging to different sensitive groups based on the rows in the normalized adjacency matrix.
    Accordingly, Spinelli et al. \cite{spinelli2021fairdrop} propose a method to modify the adjacency matrix at the training time to balance the homophily caused by the sensitive attribute. In each training iteration, they remove the edges between nodes based on a randomized response mechanism between nodes that have the same sensitive attribute value.
    \paragraph{\textbf{Learning Unbiased Embeddings}}
    The high-level idea of resolving bias for the methods in this section is to place a fairness constraint on the objective function of the representation learning model. Rahman et al. \cite{rahman2019fairwalk} try to promote fairness to Node2vec \cite{grover2016node2vec} by modifying the random walks by changing the transition probabilities to generate unbiased traces. In consequence, the generated random walk is more likely to have nodes from different groups. Khajehnejad et al. \cite{khajehnejad2022crosswalk} propose a re-weighting approach for generating the random walks, however, they assign more weights to the links that connect nodes from different groups to provide a higher chance of discovery in extreme cases that Rahman et al. \cite{rahman2019fairwalk} would have failed. Inspired by Conditional Network Embeddings \cite{kang2018conditional}, Buyl et al. \cite{buyl2020debayes} present a Bayesian approach that learns debiased representations using as strongly biased as possible prior so that the learned embeddings have minimal information about sensitive attributes in the training step.

\subsection{Summary}
In Figure \ref{fig:table_unstructured}, we summarize the papers reviewed on the identification and resolution of representation bias in unstructured data and present a side-by-side comparison between them based on different properties:

\begin{itemize}
    \item \textit{Data Type} specifies the data type targeted in the corresponding work.
    \item \textit{Task} specifies whether the algorithm identifies or resolves insufficient representation.
    \item \textit{Technique} briefly mentions the general idea of the proposed approach.
\end{itemize}

\newgeometry{left=2cm,bottom=1cm}
\begin{landscape}
\begin{figure}
    \footnotesize
    \centering
    \def\arraystretch{1.5}
    \begin{tabular}{||c|c|c|c||}\hline
    
        Paper / System&Data Type&Task&Technique\\ 
         \hline \hline  
        \pbox{8cm}{Auditing ImageNet: towards a model-driven framework for annotating demographic attributes of large-scale image datasets \cite{dulhanty2019auditing}}&
        Image&
        Identification&
        Reduction to tabular data using metadata and recognition tools to extract attributes of interest\\
        \hline 
        
        \pbox{8cm}{Gender shades: Intersectional accuracy disparities in commercial gender classification \cite{buolamwini2018gender}}&
        Image&
        Identification&
        Reduction to tabular data by annotation and counting subpopulations\\
        \hline
        
        Diversity in faces \cite{merler2019diversity}&
        Image&
        Identification&
        Reduction to tabular data by annotation and using recognition tools to extract attributes of interest\\
        \hline
        
        REVISE \cite{wang2020revise}&
        Image&
        Identification&
        Reduction to tabular data using annotated data to extract attributes of interest\\
        \hline
        
        FairFace \cite{karkkainen2021fairface}&
        Image&
        Identification&
        Distance-based analysis of the image representations in the embedding space\\
        \hline
        
        Unbiased look at data setbias \cite{torralba2011unbiased}&
        Image&
        Identification&
        Cross data set generalization by training model on a data set and testing on another one\\
        \hline
        
        Undoing the damage of data setbias \cite{khosla2012undoing}&
        Image&
        Identification&
        Learning and removing the biases in different data sets to approximate the weights for an unbiased visual world\\
        \hline
        
        Towards measuring bias in image classification \cite{schaaf2021towards}&
        Image&
        Identification&
        Identifying the bias using the attribution maps\\
        \hline
        
        Crowdsourcing detection of sampling biases in image datasets \cite{hu2020crowdsourcing}&
        Image&
        Identification&
        Crowd-sourcing approach to identify representation bias\\
        \hline
        
        Deflating data setbias using synthetic data augmentation \cite{jaipuria2020deflating}&
        Image&
        Resolution&
        Data augmentation using targeted synthetic data\\
        \hline
        
        \pbox{8cm}{Mitigating demographic bias in facial datasets with style-based multi-attribute transfer \cite{georgopoulos2021mitigating}}&
        Image&
        Resolution&
        Data augmentation w.r.t. different attributes using style transfer GANs\\
        \hline
        
        \pbox{8cm}{Exploring racial bias within face Recognition via per-subject adversarially-enabled data augmentation \cite{yucer2020exploring}}&
        Image&
        Resolution&
        Data augmentation w.r.t. race attribute using cycleGANs\\
        \hline
        
        CAMEL\cite{goel2020model}&
        Image&
        Resolution&
        Data augmentation using cycleGANs\\
        \hline
        
        REPAIR\cite{li2019repair}&
        Image&
        Resolution&
        Assigning weights instances in the data that are penalized more easily by the models\\
        \hline
        
        Measuring and mitigating unintended bias in text classification \cite{dixon2018measuring}&
        Text&
        Identification&
        \pbox{10cm}{Gender swapping and measuring the difference in evaluation score, \\Investigating the effect of disproportionate representation in the training data} \\
        \hline
        
        \pbox{8cm}{Stereotypical bias removal for hate speech detection task using knowledge-based generalizations \cite{badjatiya2019stereotypical}}&
        Text&
        Identification&
        \pbox{10cm}{Investigating the effect of disproportionate representation in the training data\\Investigating skewed occurrences across classes, Investigating skewed predicted class probability distribution} \\
        \hline
        
        \pbox{8cm}{Man is to computer programmer as woman is to homemaker? Debiasing word embeddings \cite{bolukbasi2016man}}&
        Text&
        Identification&
        \pbox{10cm}{Checking the magnitude of the dependence of a concept and two genders in the embedding space} \\
        \hline
        
        \pbox{8cm}{Black is to criminal as caucasian is to police: Detecting and removing multiclass bias in word embeddings \cite{manzini2019black}}&
        Text&
        Identification&
        \pbox{10cm}{Checking the magnitude of the dependence of a concept and multiple groups in the embedding space} \\
        \hline
        
        \pbox{8cm}{Bias in word embeddings \cite{papakyriakopoulos2020bias}}&
        Text&
        Identification&
        \pbox{12cm}{Checking the magnitude of the dependence of a concept and two groups in the embedding space} \\
        \hline
        
        \pbox{8cm}{Gender and representation bias in GPT-3 generated stories \cite{lucy2021gender}}&
        Text&
        Resolution&
        \pbox{12cm}{Prompt design to reduce gender bias in text generation} \\
        \hline
        
        \pbox{8cm}{Learning gender-neutral word embeddings \cite{zhao2018gender}}&
        Text&
        Resolution&
        \pbox{12cm}{Augmenting corpora by appending the gender-swapped version of the text} \\
        \hline
        
        \pbox{8cm}{Getting gender right in neural machine translation \cite{vanmassenhove2019getting}}&
        Text&
        Resolution&
        \pbox{12cm}{Augmenting corpora by gender-tagging} \\
        \hline
        
        \pbox{8cm}{Reducing gender bias in abusive language detection \cite{park2018reducing}}&
        Text&
        Resolution&
        \pbox{12cm}{Transfer learning
        from an already bias-free data set and fine-tune on the biased data to train a model} \\
        \hline
        
        \pbox{8cm}{Man is to computer programmer as woman is to homemaker? Debiasing word embeddings \cite{bolukbasi2016man}}&
        Text&
        Resolution&
        \pbox{10cm}{Eliminating the gender-pair associations from gender-neutral words by making it orthogonal to the gender vector in the embedding space} \\
        \hline
        
        \pbox{8cm}{Black is to criminal as caucasian is to police: Detecting and removing multiclass bias in word embeddings \cite{manzini2019black}}&
        Text&
        Resolution&
        \pbox{10cm}{Eliminating the non-binary gender/race pair associations from gender/race neutral words by making it orthogonal to the gender/race vector in the embedding space} \\
        \hline
        
        \pbox{8cm}{Toward understanding bias correlations for mitigation in NLP \cite{cheng2022toward}}&
        Text&
        Resolution&
        \pbox{10cm}{Joint bias removal w.r.t. to different sensitive attributes by neutralizing the word vectors in the embedding space} \\
        \hline
        
        \pbox{8cm}{Learning gender-neutral word embeddings \cite{zhao2018learning}}&
        Text&
        Resolution&
        \pbox{10cm}{Separating information about the sensitive attribute by keeping it in another dimension} \\
        \hline
        
        \pbox{8cm}{Quantifying bias in automatic speech recognition \cite{feng2021quantifying}, Racial disparities in automated speech recognition \cite{koenecke2020racial}, Towards Measuring Fairness in speech recognition: casual conversations data set transcriptions \cite{liu2022towards} }&
        Speech&
        Identification&
        \pbox{10cm}{Measuring error rate whiting subgroups identified through annotation or automatic recognition tools} \\
        \hline
        
        \pbox{8cm}{Debayes: a bayesian method for debiasing network embeddings \cite{buyl2020debayes}}&
        Graph&
        Identification&
        \pbox{10cm}{Using mixing coefficient to evaluate the homophily of a graph w.r.t. a specific attribute} \\
        \hline
        
        \pbox{8cm}{Compositional fairness constraints for graph embeddings \cite{bose2019compositional}}&
        Graph&
        Identification&
        \pbox{10cm}{Identifying bias in node embedding level using the notion of Representation Bias} \\
        \hline
        
        \pbox{8cm}{All of the fairness for edge prediction with optimal transport \cite{laclau2021all}}&
        Graph&
        Resolution&
        \pbox{10cm}{Repair graph by aligning node distributions of nodes belonging to different sensitive groups} \\
        \hline
        
        \pbox{8cm}{Fairdrop: Biased edge dropout for enhancing fairness in graph representation learning \cite{spinelli2021fairdrop}}&
        Graph&
        Resolution&
        \pbox{10cm}{Modifying the adjacency matrix at the training time to balance the homophily caused by the sensitive attribute} \\
        \hline
        
        \pbox{8cm}{Debayes: a bayesian method for debiasing network embeddings \cite{buyl2020debayes}}&
        Graph&
        Resolution&
        \pbox{10cm}{Bayesian approach to learn debiased representations using a strongly biased prior} \\
        \hline
        
        \pbox{8cm}{CrossWalk: fairness-enhanced node representation learning \cite{khajehnejad2022crosswalk}}&
        Graph&
        Resolution&
        \pbox{10cm}{Re-weighting approach by assigning more weights to the links that connect nodes from different groups for generating random walks } \\
        \hline
        
        \pbox{8cm}{Fairwalk: Towards fair graph embedding \cite{rahman2019fairwalk}}&
        Graph&
        Resolution&
        \pbox{10cm}{Modifying the random walks through changing the transition probabilities to generate unbiased traces} \\
        \hline
        
    \end{tabular}
    \caption{Properties of different techniques for identifying and resolving representation bias in unstructured data.}
    \label{fig:table_unstructured}
\end{figure}
\end{landscape}
\restoregeometry
\section{Conclusion}\label{sec:conclusion}
In this paper, we surveyed techniques for the identification and resolution of representation bias in data. 
After reviewing the fairness literature at a high level, we provided a thorough overview of the problem definition, the causes, and how to measure and quantify this phenomenon in both structured and unstructured data. Depending on the data type, we then presented taxonomies based on multiple dimensions and had side-by-side comparisons of the techniques. We discussed the details of several algorithms to illustrate the different challenges and the problems they address. 
Two promising research directions we envision being important are:
\begin{itemize}
    \item \textit{Addressing representation bias in other types of data sets.} As we discussed in section \ref{sec:unstructured}, with the extension of the problem scope to new data types such as streaming data, spatio-temporal data, etc., new challenges arise and the current solutions may not be directly extendable. 
    \item \textit{More metrics for measuring representation bias.} Existing works have introduced \textit{coverage} and \textit{representation rate} for measuring representation bias. However, each metric has potential shortcomings that provide new research opportunities. Furthermore, when it comes to data quality and trust measures in data, there is no such thing as ``enough'' and there is always room for improvement. 
\end{itemize} 

\section*{Acknowledgements}

This research was supported in part by the National Science Foundation, under grants 2107290, 1741022, 1934565, and 2106176.
\bibliographystyle{ACM-Reference-Format}
\bibliography{ref}
\end{document}